\documentstyle[12pt]{article}
\newcommand{\seqnoll}{\setcounter{equation}{0}}

\newcommand{\eq}[1]{\mbox{Eq.~(\ref{#1})}}
\def\figcap#1{\refstepcounter{figure}\par\small\bf Figure \thefigure: \sl#1}

\newcommand{\bort}[1]{}
\def\id{\leavevmode\hbox{\small1\kern-3.3pt\normalsize1}}
\newcommand{\goto}{\rightarrow}

\newcommand{\fig}[1]{\mbox{Fig.~\ref{#1}}}
\newcommand{\non}{\nonumber}
\newcommand{\nn}{\nonumber\\}
\newcommand{\abs}[1]{\left|#1\right|}

\newcommand{\inv}[1]{\frac{1}{#1}}
\newcommand{\cF}{{\cal F}}

\def\simleq{\; \raise0.3ex\hbox{$<$\kern-0.75em
      \raise-1.1ex\hbox{$\sim$}}\; }
\def\simgeq{\; \raise0.3ex\hbox{$>$\kern-0.75em
      \raise-1.1ex\hbox{$\sim$}}\; }
\def\void{}
\def\labelmark{}

\newenvironment{formula}[1]%
{\def\labelname{#1}
\ifx\void\labelname\begin{equation}
\else\labelmark\begin{equation}\label{\labelname}\fi}%
{\ifx\void\labelname\end{equation}\else\end{equation}\fi}

\newenvironment{formulas}[1]%
{\def\labelname{#1}
\ifx\void\labelname\begin{equation}
\else\labelmark\begin{equation}\label{\labelname}\fi
\begin{array}{lllllll}}%
{\end{array}\ifx\void\labelname\end{equation}\else
\end{equation}\fi}

\def\erf{{\rm erf}}
\def\O{{\cal O}}
\def\.{~~.}
\def\simleq{\; \raise0.3ex\hbox{$<$\kern-0.75em
      \raise-1.1ex\hbox{$\sim$}}\; }
\def\simgeq{\; \raise0.3ex\hbox{$>$\kern-0.75em
      \raise-1.1ex\hbox{$\sim$}}\; }

\def\noi{\noindent}
\def\ie{{\rm i.e.}}
\def\apriori{{\it a priori}}

\def\({\left(}
\def\){\right)}

\def\>{\rangle}
\def\<{\langle}

\def\Tr{{\rm Tr}}

\def\P{{\cal P}}
\def\n{{\hat n}}

\def\bfor#1{\begin{formula}{#1}}
\def\efor{\end{formula}}

\begin{document}
\vspace*{-2cm}
\begin{flushright}
\vbox{\noindent
CERN/TH 95-333\\
G\"oteborg ITP 95-27\\
atom-ph/9601004\\
Christmas 1995}
\end{flushright}
\vspace{1cm}
\begin{center}
{\Huge\bf Dynamics, Correlations and\\[3mm] Phases of the Micromaser}\\
\vspace{10mm}
\noi
{\large Per Elmfors\footnote{email: elmfors@cern.ch}$^{,a)}$,
Benny Lautrup\footnote{email: lautrup@connect.nbi.dk}$^{,a,b)}$
and  Bo-Sture Skagerstam\footnote{email:
tfebss@fy.chalmers.se}$^{,c,d)}$\\[10mm]
}
{
\small
$^{a)}$ CERN,   TH-Division, CH-1211 Geneva 23, Switzerland \\[1mm]
\small
$^{b)}${\sc connect}, The Niels Bohr Institute, Blegdamsvej 17,\\
\small        DK-2100 Copenhagen, Denmark \\[1mm]
\small $^{c)}$Institute of Theoretical Physics, Chalmers University of
 Technology \\
\small and G\"oteborg University, S-412 96 G\"oteborg, Sweden\\[1mm]
\small $^{d)}$Department of Physics, University of Trondheim,
N-7055 Dragvoll, Norway\\
}
\end{center}
{\abstract{
The micromaser possesses a variety of dynamical phase transitions  parametrized
by the flux of atoms and the time-of-flight of the atom within the  cavity.  We
discuss how these phases may be revealed to  an  observer  outside  the  cavity
using the long-time correlation length in the atomic beam. Some  of  the  phase
transitions are not reflected in the average excitation level of  the  outgoing
atom, which is the commonly used observable. The correlation length is directly
related to the leading eigenvalue of the  time  evolution  operator,  which  we
study in order to elucidate the phase structure. We find that as a function  of
the time-of-flight the transition from  the  thermal  to  the  maser  phase  is
characterized  by  a  sharp  peak  in  the  correlation  length.   For   longer
times-of-flight there is a transition to a phase where the  correlation  length
grows exponentially  with  the  flux.  We  present  a  detailed  numerical  and
analytical treatment of the different phases and  discuss  the  physics  behind
them. }}
\vfill
{{\vbox{\noindent
CERN/TH 95-333\\
G\"oteborg ITP 95-27\\
atom-ph/9601004\\
Christmas 1995}}}
\thispagestyle{empty}
\newpage
\vspace*{-3cm}
\tableofcontents
\thispagestyle{empty}
\newpage
\setcounter{page}{1}
\setcounter{footnote}{0}
\section{Introduction}
\seqnoll
The  highly  idealized  physical  system  of  a  single  two-level  atom  in  a
superconducting   cavity,   interacting   with    a    quantized    single-mode
electromagnetic field, has  been  experimentally  realized  in  the  micromaser
\cite{Goy83}--\cite{Walther95} and microlaser systems \cite{An94}. Details  and
references  to  the   literature   can   be   found   in   e.g.   the   reviews
\cite{Haroche92}--\cite{Milonni91}. In the absence of dissipation (and  in  the
rotating wave approximation) the two-level atom and its  interaction  with  the
radiation field is well  described  by  the
Jaynes--Cummings  (JC)  Hamiltonian
\cite{Jaynes63}. Since  this  model  is  exactly  solvable  it  has  played  an
important role in the development  of  modern  quantum  optics  (for  a  recent
account see e.g. Refs.~\cite{Barnett86,Milonni91}). The  JC model
predicts  non-classical
phenomena, such as revivals of the initial excited   state   of   the   atom
\cite{Meystre74}--\cite{Arroyo90}, experimental signs  of   which
have been reported \cite{Rempe87}.

Correlation  phenomena  are  important  ingredients  in  the  experimental  and
theoretical investigation of physical systems. Intensity correlations of  light
was e.g. used by Hanbury--Brown
and Twiss \cite{HBT56} as a tool to determine  the
angular diameter of distant stars. The quantum theory of intensity correlations
of light was later developed by Glauber \cite{Glauber63}.  These  methods  have
a wide range of physical applications including investigation of  the
space-time
evolution    of    high-energy     particle     and     nuclei     interactions
\cite{Boal90,Skagerstam94}. In the case of  the  micromaser  we  have  recently
suggested \cite{Elmforsetal95} that correlation measurements on  atoms  leaving
the micromaser system can be used to infer properties of the quantum state  of
the
radiation field in the cavity.

In this  paper  we  present  a  detailed  account  of  the  role  of  long-time
correlations in the outgoing atomic beam and  their  relation  to  the  various
phases of the micromaser system. Fluctuations in the number  of  atoms  in  the
lower maser level for a fixed transit time $\tau $ is known to  be  related  to
the photon-number  statistics  \cite{Filipowicz86}--\cite{PaulRichter91}.  The
experimental  results  of  \cite{Rempe90}  are  clearly  consistent  with   the
appearance of non-classical, sub-Poissonian statistics of the radiation  field,
and exhibit the intricate correlation between the atomic beam and  the  quantum
state of the cavity. Related work on characteristic statistical  properties  of
the beam of  atoms  emerging  from  the  micromaser  cavity  may  be  found  in
Ref.~\cite{Brigeletal94}
\cite{WagnerSW94,Herzog94}.

The paper is organized  as  follows.  In  Section~\ref{basic}  we  discuss  the
standard theoretical framework  for  the  micromaser  and  introduce  some  new
notation.  A  general  discussion  of  long-time  correlations  is   given   in
Section~\ref{correlations}, where  we  also  determine  the  correlation
length
numerically. Before entering the analytic investigation of the phase  structure
we introduce some useful concepts in  Section~\ref{analytic}  and  discuss  the
eigenvalue problem for the correlation length. The heart of the paper  lies  in
Section~\ref{Phasestructure},
where  details  of  the  different   phases   are
analysed. In Section~\ref{Spread} we study effects related to the finite spread
in atomic velocities. The phase boundaries are defined in the limit of an
infinite
flux of atoms, but there are several interesting effects related to finite
fluxes
as well. We discuss these issues in Section~\ref{finite}. Finally we  summarize
our results in Section~\ref{conclusions}.


\section{Basic micromaser theory}
\label{basic}
\seqnoll
In the micromaser a beam of excited atoms is sent through  a  cavity  and  each
atom interacts with the cavity during a well-defined transit time  $\tau$.  The
theory     of     the      micromaser      has      been      developed      in
\cite{Filipowicz86,Filipowicz86a}, and in this section we  briefly  review  the
standard theory, generally following the notation of that paper. We assume that
excited atoms are injected into the cavity at an average rate $R$ and that  the
typical decay rate for photons in the cavity is $\gamma$. The number  of  atoms
passing the cavity  in  a  single  decay  time  $N=R/\gamma$  is  an  important
dimensionless parameter, effectively controlling the average number of  photons
stored in a high-quality cavity. We shall assume that the time $\tau$
during which the  atom
interacts with the cavity is so small that effectively only one atom  is  found
in the cavity at any  time,  \ie\  $R\tau\ll1$.  A  further  simplification  is
introduced by assuming that the cavity decay time  $1/\gamma$  is  much  longer
than the interaction time, \ie\ $\gamma\tau\ll1$, so that damping  effects  may
be ignored while the atom passes through the  cavity.  This  point  is  further
elucidated  in  Appendix~\ref{AppDamping}.  In  the   typical   experiment   of
Ref.~\cite{Rempe90}   these   quantities   are   given   the   values   $N=10$,
$R\tau=0.0025$ and $\gamma\tau=0.00025$.


\subsection{The Jaynes--Cummings Model}

The electromagnetic interaction between a two-level atom with level  separation
$\omega_0$ and a single mode with frequency $\omega$ of the radiation field  in
a cavity is described,
in the rotating wave approximation, by the Jaynes--Cummings
(JC) Hamiltonian \cite{Jaynes63}

\begin{formula}{JCH}
H=\omega a^{*} a+\frac12\omega_0\sigma_z+g(a\sigma_++a^*\sigma_-)~~,
\end{formula}

\noi where the coupling constant $g$  is  proportional  to  the  dipole  matrix
element of the atomic transition\footnote{This coupling constant turns  out  to
be identical to the single photon Rabi frequency  for  the  case  of  vanishing
detuning, \ie\ $g=\Omega$. There is actually some confusion in  the  literature
about what is called the Rabi frequency \cite{Knight80}.  With  our
definition,
the  energy  separation  between   the   shifted   states   at   resonance   is
$2\Omega$.}. We use the Pauli matrices to describe the  two-level  atom  and
the   notation   $\sigma_\pm=(\sigma_x\pm   i\sigma_y)/2$.   For   $g=0$    the
atom-plus-field  states  $|n,s\>$  are  characterized  by  the  quantum  number
$n=0,1,\ldots$ of  the  oscillator  and  $s=\pm$  for  the  atomic  levels
(with $-$ denoting the ground state).  At
resonance $\omega=\omega_0$ the levels $|n-1,+\>$ and $|n,-\>$  are  degenerate
for $n\ge1$ (excepting the ground state $n=0$), but this degeneracy  is  lifted
by  the  interaction.  For  arbitrary  coupling  $g$  and  detuning   parameter
$\Delta\omega= \omega_0-\omega$ the system reduces to a  $2\times2$  eigenvalue
problem,
which may be trivially solved. The result is that two  new  levels  are
formed as superpositions of the previously degenerate ones with a separation in
energy  $E_{n-1,+}-E_{n,-}=\sqrt{\Delta\omega^2+4g^2n}$.  The  system  performs
Rabi oscillations with this frequency between the original, unperturbed  states
with transition probabilities \cite{Jaynes63}
\begin{formulas}{TransEl}
|\<n,-|e^{-iH \tau}|n,-\>|^2&=&1-q_n(\tau)~~,\\
|\<n-1,+|e^{-iH \tau}|n,-\>|^2&=&q_n(\tau)~~,\\
|\<n,+|e^{-iH \tau}|n,+\>|^2&=&1-q_{n+1}(\tau)~~,\\
|\<n+1,-|e^{-iH \tau}|n,+\>|^2&=&q_{n+1}(\tau)~~.\\
\end{formulas}

\noi These are all expressed in terms of

\begin{formula}{}
q_n(\tau)=\frac{g^2n}{g^2n+\frac14\Delta\omega^2}
\sin^2\(\tau\sqrt{g^2n+{\textstyle\frac14}\Delta\omega^2}\) ~~.
\end{formula}

\noi Notice that for $\Delta\omega=0$ we have
$q_n=\sin^2 (g\tau\sqrt n )$.
Most of the following discussion will be limited to this case.

Denoting the probability of finding $n$ photons in the cavity by $p_n$ we  find
the conditional probability that an excited atom decays to the ground state  in
the cavity to be

\begin{formula}{}
\P(-)=\<q_{n+1}\>=\sum_{n=0}^\infty q_{n+1} p_n ~~.
\end{formula}

\noi It is this sum over the incommensurable frequencies $g\sqrt n$  that  is
the cause of some of the most important properties of the micromaser, such as
quantum collapse and revivals (see e.g. Refs.
\cite{Averbukh89a}--\cite{Fleischhauer93}). These  effects  are  most
easily displayed in the case that the cavity field is coherent  with  Poisson
distribution

\begin{formula}{Poisson}
p_n=\frac{\<n\>^n}{n!}e^{-\<n\>} ~~.
\end{formula}

\noi In the more realistic case, where the changes of the cavity field due to
the passing atoms is taken into account, a complicated statistical  state  of
the     cavity     arises     \cite{Filipowicz86},
\cite{Wright89}--\cite{Herzog95}
\bort{\cite{Filipowicz86,Wright89,Guzman89,Herzog95}}
\bort{
(see Fig.~\ref{FigLandscape})
}
\cite{BogarBH94}. It is the  details  of  this  state  that  are
investigated in this paper.


\subsection{Mixed states}
\label{mixed}

The above formalism is directly applicable when  the  atom  and  the  radiation
field are both in pure states initially. In general the  statistical  state  of
the system is described by an  initial  density  matrix  $\rho$,  which
evolves
according to the usual rule $\rho\to\rho(t)=\exp(-iHt)\rho\exp(iHt)$.
If we disregard, for
the moment, the decay of the cavity field due to interactions with the
environment, the evolution is governed by the JC Hamiltonian in \eq{JCH}. It is
natural to assume that the atom and the radiation field of the cavity initially
are completely uncorrelated so that the initial density matrix factorizes in  a
cavity part and a product of $k$ atoms as

\begin{formula}{rhofact}
    \rho=\rho_C\otimes\rho_{A_1}\otimes\rho_{A_2}\otimes \cdots \otimes
\rho_{A_k}~~.
\end{formula}

\noi When the first atom $A_1$ has passed through the cavity,  part  of  this
factorizability is destroyed by the interaction and the state has become

\begin{formula}{rhotau}
\rho(\tau)=\rho_{C,A_1}(\tau)   \otimes\rho_{A_2}\otimes \cdots \otimes
\rho_{A_k}~~.
\end{formula}

\noi   The   explicit   form   of   the   cavity-plus-atom   entangled    state
$\rho_{C,A_1}(\tau)$  is  analysed  in  Appendix~\ref{AppDamping}.  After   the
interaction, the cavity decays, more atoms pass through and the  state  becomes
more and more entangled. If we decide never  to  measure  the  state  of  atoms
$A_1\ldots  A_i$  with  $i<k$,  we  should  calculate  the   trace   over   the
corresponding states and only the $\rho_0$-component remains.  Since  the  time
evolution  is  linear,  each  of  the   components   in   \eq{rhotau}   evolves
independently, and it does not matter when we calculate the trace. We can do
it
after each atom has passed the cavity, or at the end  of  the  experiment.  For
this we do not even have to assume that the  atoms  are  non-interacting  after
they leave the cavity, even though this simplifies the time evolution. If we do
perform a measurement of the state of an intermediate atom $A_i$, a correlation
can be observed between that result and a measurement of atom  $A_k$,  but  the
statistics of the unconditional measurement of  $A_k$  is  not  affected  by  a
measurement of $A_i$. In a real experiment also the efficiency of the measuring
apparatus should be taken into account when using  the  measured  results  from
atoms  $A_1,\ldots,A_i$  to  predict  the  probability  of  the  outcome  of  a
measurement of $A_k$ (see Ref. \cite{Brigeletal94} for a detailed investigation
of this case).

As a generic case let us assume that the initial  state  of  the  atom  is  a
diagonal mixture of excited and unexcited states

\begin{formula}{InitAtom}
\rho_A=\(\begin{array}{cc}a&0\\0&b\\\end{array}\)~~,
\end{formula}

\noi where, of course, $a,b\ge0$ and $a+b=1$. Using that both  preparation  and
observation are diagonal in the atomic states, it may  now  be  seen  from  the
transition elements in \eq{TransEl} that  the  time  evolution  of  the  cavity
density matrix does not mix different diagonals of this matrix.  Each  diagonal
so to speak ``lives its own life'' with respect to dynamics. This implies  that
if the initial cavity density matrix is diagonal, \ie\ of the form

\begin{formula}{}
\rho_C=\sum_{n=0}^\infty p_n |n\>\<n|~~,
\end{formula}

\noi with $p_n\ge0$ and  $\sum_{n=0}^\infty  p_n=1$,  then  it  stays  diagonal
during the interaction between atom and cavity and may always be described by a
probability distribution $p_n(t)$. In fact,  we  easily  find  that  after  the
interaction we have

\begin{formula}{Master}
p_n(\tau)=a q_n(\tau)p_{n-1}+bq_{n+1}(\tau)p_{n+1}
+(1-a q_{n+1}(\tau)-bq_{n}(\tau))p_n~~,
\end{formula}

\noi where the first term is the probability of decay for  the  excited  atomic
state, the second the probability of excitation for the  atomic  ground  state,
and the third is the probability  that  the  atom  is  left  unchanged  by  the
interaction. It is convenient to write this in matrix form
\cite{Herzog94}

\begin{formula}{}
p(\tau)=M(\tau)p~~,
\end{formula}

\noi with a transition matrix $M=M(+)+M(-)$ composed of two parts, representing
that the outgoing atom is either in the excited state (+) or  in  the  ground
state ($-$). Explicitly we have

\begin{formulas}{Pump}
M(+)_{nm}=&bq_{n+1}\delta_{n+1,m}
+a(1-q_{n+1})\delta_{n,m}~~,\\
M(-)_{nm}=&aq_n\delta_{n,m+1}
+b(1-q_{n})\delta_{n,m}~~.
\end{formulas}

\noi Notice that these formulas are completely classical and may  be  simulated
with a standard Markov  process.  The  statistical  properties  are  not
quantum
mechanical as long as the incoming atoms have a diagonal density matrix and  we
only measure elements in the diagonal. The only quantum mechanical  feature  at
this stage is the discreteness  of  the  photon  states,  which  has  important
consequences for the correlation  length  (see Section~\ref{LLCav}).  If  the
atomic
density matrix has off-diagonal elements, the above formalism breaks down.  The
reduced cavity density matrix will then  also  develop  off-diagonal  elements,
even if initially it is diagonal. We shall not go further  into  this  question
here (see for example Refs.~\cite{Krause86}--\cite{Zaheer89}).
\bort{\cite{Krause86,Lugiato87,Zaheer89}}


\subsection{The lossless cavity}\label{LLCav}

The above discrete master equation (\ref{Master}) describes the pumping of a
lossless cavity with a beam of atoms. After $k$ atoms have  passed  through
the cavity, its state has become  $M^kp$.  In  order  to  see  whether  this
process may reach statistical  equilibrium  for  $k\to\infty$  we  write
Eq.~(\ref{Master}) in the form

\begin{formula}{Current}
p_n(\tau)=p_n+J_{n+1}-J_n~~,
\end{formula}

\noi where $J_n=-aq_np_{n-1}+bq_np_n$. In statistical equilibrium we must  have
$J_{n+1}=J_n$, and the common value $J=J_n$ for all $n$ can only be zero  since
$p_n$, and therefore $J$, has to vanish for $n\to\infty$. It  follows  that
this
can only be the case for $a<b$ \ie\ $a<0.5$. There must thus be fewer than 50\%
excited atoms in the beam, otherwise the lossless cavity blows up. If $a<0.5$,
the cavity will reach an equilibrium distribution of  the  form  of  a  thermal
distribution  for   an   oscillator   $p_n=(1-a/b)(a/b)^n$.   The   statistical
equilibrium may be shown to be stable, \ie\ that all non-trivial eigenvalues of
the matrix $M$ are real and smaller than 1.


\subsection{The dissipative cavity}

A single oscillator interacting with an environment having a huge number of
degrees of freedom, for example a heat bath, dissipates energy
according to the well-known damping formula (see for example
\cite{Agarwal73,Walls95}):

\begin{formulas}{Damping}
{\displaystyle\frac{d\rho_C}{dt}}=&~~i[\rho_C,\omega a^*a]\\
&-\frac12\gamma(n_b+1)(a^*a\rho_C+\rho_C a^*a-2a\rho_C a^*)\\
&-\frac12\gamma n_b(aa^*\rho_C+\rho_C aa^*-2a^*\rho_C a)~~,\\
\end{formulas}

\noi where $n_b$  is  the  average  environment  occupation  number  at  the
oscillator frequency and $\gamma$ is the decay constant. This evolution also
conserves diagonality, so we have for any diagonal cavity state:

\begin{formula}{Dissipative}
\frac1\gamma\frac{dp_n}{dt}=-(n_b+1)(np_n-(n+1)p_{n+1})-n_b((n+1)p_n-np_{n-1})
{}~~,
\end{formula}

\noi which of course conserves probability. The right-hand  side  may  as  for
Eq.~(\ref{Current}) be written as $J_{n+1}-J_n$ with
$J_n=(n_b+1)np_n-n_bnp_{n-1}$
and the same arguments as above lead to a thermal equilibrium distribution with

\begin{formula}{Thermal}
p_n=\frac1{1+n_b}\(\frac{n_b}{1+n_b}\)^n ~~.
\end{formula}


\subsection{The discrete master equation}

We now take into account both pumping and damping. Let the next atom arrive  in
the cavity after a time $T\gg\tau$. During this interval the cavity damping  is
described by Eq.~(\ref{Dissipative}), which we shall write in the form

\begin{formula}{}
\frac{dp}{dt}=-\gamma  L_C  p~~,
\end{formula}

\noi where $L_C$ is the cavity decay matrix from above

\begin{formula}{LC}
(L_C)_{nm}=(n_b+1)(n\delta_{n,m}-(n+1)\delta_{n+1,m})
+n_b((n+1)\delta_{n,m}-n\delta_{n-1,m}) ~~.
\end{formula}

\noi The statistical state of the cavity when the next atom arrives is  thus
given by

\begin{formula}{stat1}
p(T)=e^{-\gamma L_CT}M(\tau)p ~~.
\end{formula}

\noi
In using the full interval $T$ and not $T-\tau$ we allow for
the decay  of
the cavity in the interaction time, although this decay is not properly
included  with  the  atomic  interaction  (for  a   correct   treatment   see
Appendix~\ref{AppDamping}).

This would be the master equation describing the evolution of the cavity if the
atoms in the beam arrived with definite and known intervals. More commonly, the
time  intervals  $T$  between  atoms  are  Poisson-distributed   according   to
$d\P(T)=\exp(-RT)RdT$  with  an  average  time  interval  $1/R$  between  them.
Averaging the exponential in Eq.~(\ref{stat1}) we get

\begin{formula}{aveL}
\<p(T)\>_T= S p~~,
\end{formula}

\noi where

\begin{formula}{Discrete1}
S=\frac1{1+L_C/N}M~~,
\end{formula}

\noi and $N=R/\gamma$ is the dimensionless  pumping  rate  already  introduced.

Implicit in the above consideration is the lack  of  knowledge  of  the  actual
value of the atomic state after the interaction. If we know that the  state  of
the atom is $s=\pm$ after the  interaction,  then  the  average  operator  that
transforms the cavity state is  instead

\begin{formula}{Discrete2}
S(s)=(1+L_C/N)^{-1}M(s)~,
\end{formula}

\noi  with  $M(s)$ given by Eq.~(\ref{Pump}).

Repeating the process for a sequence of $k$ unobserved atoms we find  that  the
initial probability distribution $p$ becomes $S^kp$. In the general  case  this
Markov process converges towards a  statistical  equilibrium  state  satisfying
$Sp=p$, which has the solution \cite{Filipowicz86,Lugiato87} for $n\ge1$

\begin{formula}{Equilibrium}
p_n=p_0\prod_{m=1}^n \frac{n_bm+Naq_m}{(1+n_b)m+Nbq_m}~~.
\end{formula}

\noi The overall constant $p_0$ is determined by $\sum_{n=0}^\infty p_n=1$.


\section{Correlations}
\label{correlations}
\seqnoll

\begin{figure}[htb]
\unitlength=1mm
\begin{picture}(140,100)(0,0)
\includegraphics{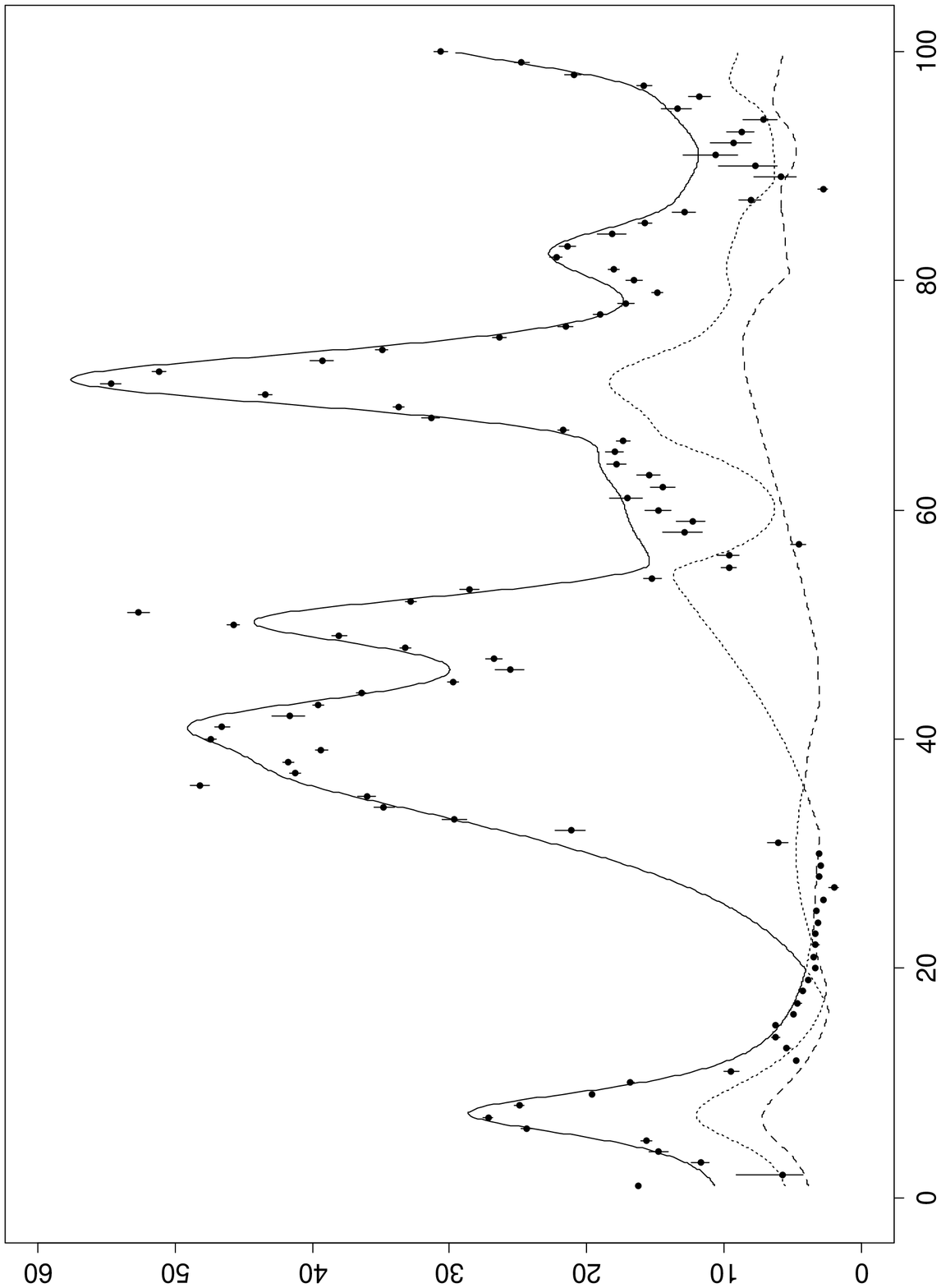}
\put(75,5){\small$\tau~[\mu {\rm s}]$}
\put(0,53){\small$R\xi$}
\put(20,90){\small$^{85}$Rb 63p$_{3/2}\leftrightarrow$ 61d$_{5/2}$}
\put(20,80){\small$R=50~{\rm s}^{-1}$}
\put(20,70){$n_b=0.15$}
\put(20,60){$\gamma=5~{\rm s}^{-1}$}
\end{picture}
\figcap{Comparison of theory  (solid  curve)  and  MC  data  (dots)  for  the
correlation length $R\xi$ (sample size $10^6$ atoms). The dotted  and  dashed
curves correspond to subleading eigenvalues ($\kappa_{2,3}$) of  the  matrix
$S$. The parameters are those of the experiment in Ref.~\cite{Rempe90}. }
\label{FigXi}
\end{figure}

After studying stationary single-time properties of the micromaser,  such  as
the average photon number in the cavity and the  average  excitation  of  the
outgoing atoms, we now proceed to dynamical properties. Correlations  between
outgoing atoms are not only determined by the equilibrium distribution in the
cavity but also by its approach to this equilibrium. Short-time  correlations,
such as the correlation between two  consecutive atoms
\cite{PaulRichter91,Herzog94},
are difficult to determine experimentally,  because  they  require  efficient
observation of the states of atoms emerging from  the  cavity in
rapid  succession.
We propose instead to study and  measure  long-time  correlations,
which  do  not  impose  the  same  strict  experimental   conditions.   These
correlations turn out  to  have  a  surprisingly  rich  structure  (see  Fig.
\ref{FigXi}) and reflect global properties of  the  photon  distribution.  In
this section we introduce the concept of long-time correlations  and  present
two ways of calculating them numerically. In the following sections we  study
the analytic properties of these correlations and elucidate their relation to
the dynamical phase structure, especially those aspects that are poorly
seen in the single-time observables or short-time correlations.


\subsection{Atomic beam observables}

Let us imagine that we know the state of all the  atoms  as  they  enter  the
cavity, for example that they are all  excited,  and  that  we  are  able  to
determine the state of each atom as it exits from the cavity. We shall assume
that the initial beam is statistically stationary, described by  the  density
matrix (\ref{InitAtom}), and that we have obtained an experimental record  of
the exit states of all the atoms after the  cavity  has  reached  statistical
equilibrium with the beam.
The effect of non-perfect measuring efficiency
has been considered in several papers \cite{Brigeletal94,WagnerSW94,Herzog94}
but we ignore that complication since it is a purely experimental problem.
{}From this record  we  may  estimate  a  number  of
quantities, for example the probability  of  finding  the  atom  in  a  state
$s=\pm$ after the interaction, where we choose $+$ to represent  the  excited
state and $-$ the ground state. The  probability  may  be  expressed  in  the
matrix form

\begin{formula}{SingleP}
\P(s)={u^0}^\top M(s) p^0~~,
\end{formula}

\begin{figure}[p]
\unitlength=1mm
\begin{picture}(140,80)(0,0)
\includegraphics{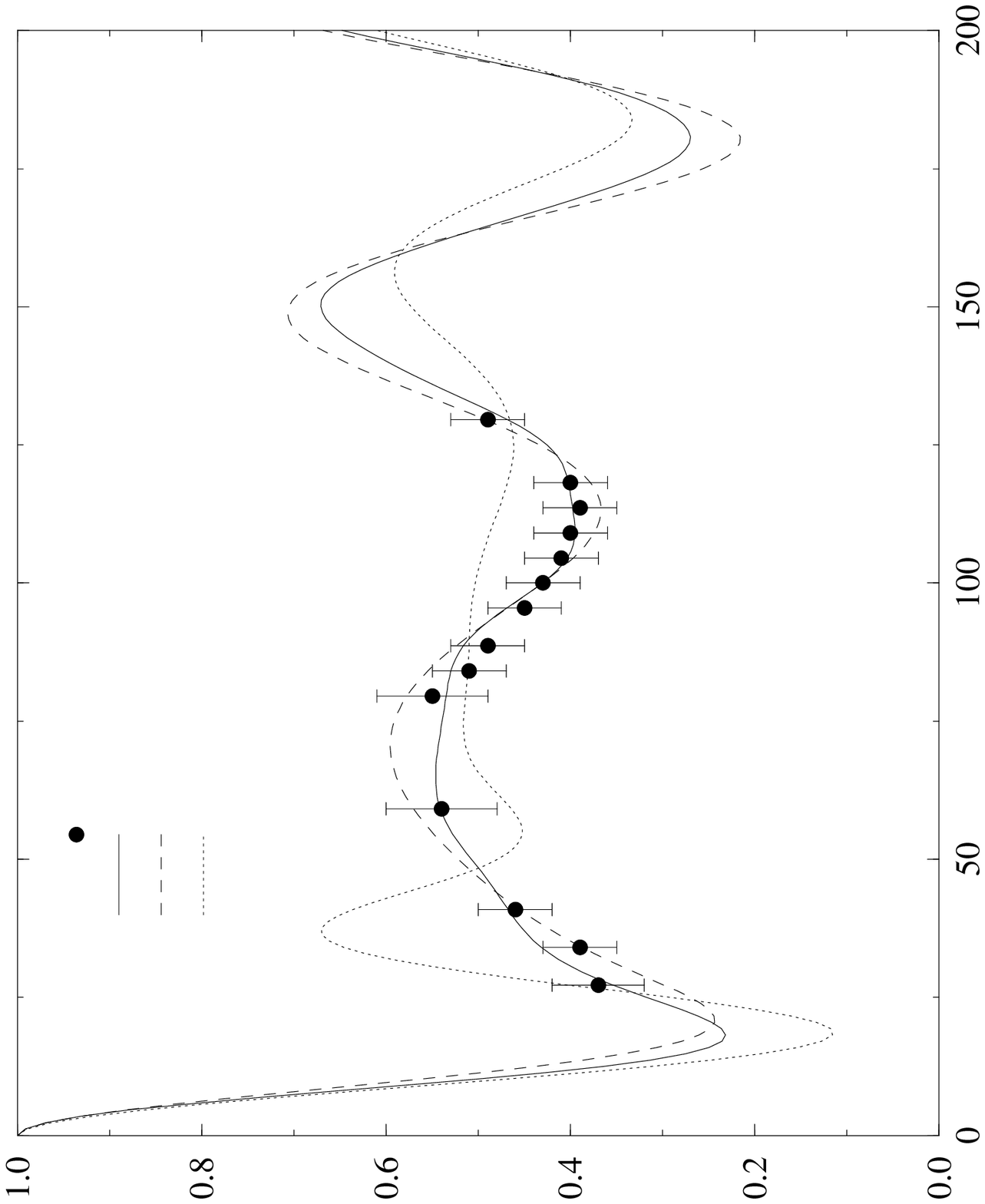}
\put(52,87){\small Experiment}
\put(52,83){\small Equilibrium: $n_b=2,~N=1$}
\put(52,79){\small Thermal: $n_b=2$}
\put(52,75){\small Poisson: $\<n\>=2.5$}
\put(70,0){\small$\tau~[\mu {\rm s}]$}
\put(0,50){\small$\P(+)$}
\put(60,23){\small$^{85}$Rb 63p$_{3/2}\leftrightarrow$ 61d$_{5/2}$}
\put(60,13){\small$R=500~{\rm s}^{-1}$,~$n_b=2$,~$\gamma=500~{\rm s}^{-1}$}
\end{picture}

\begin{picture}(140,100)(0,0)
\includegraphics{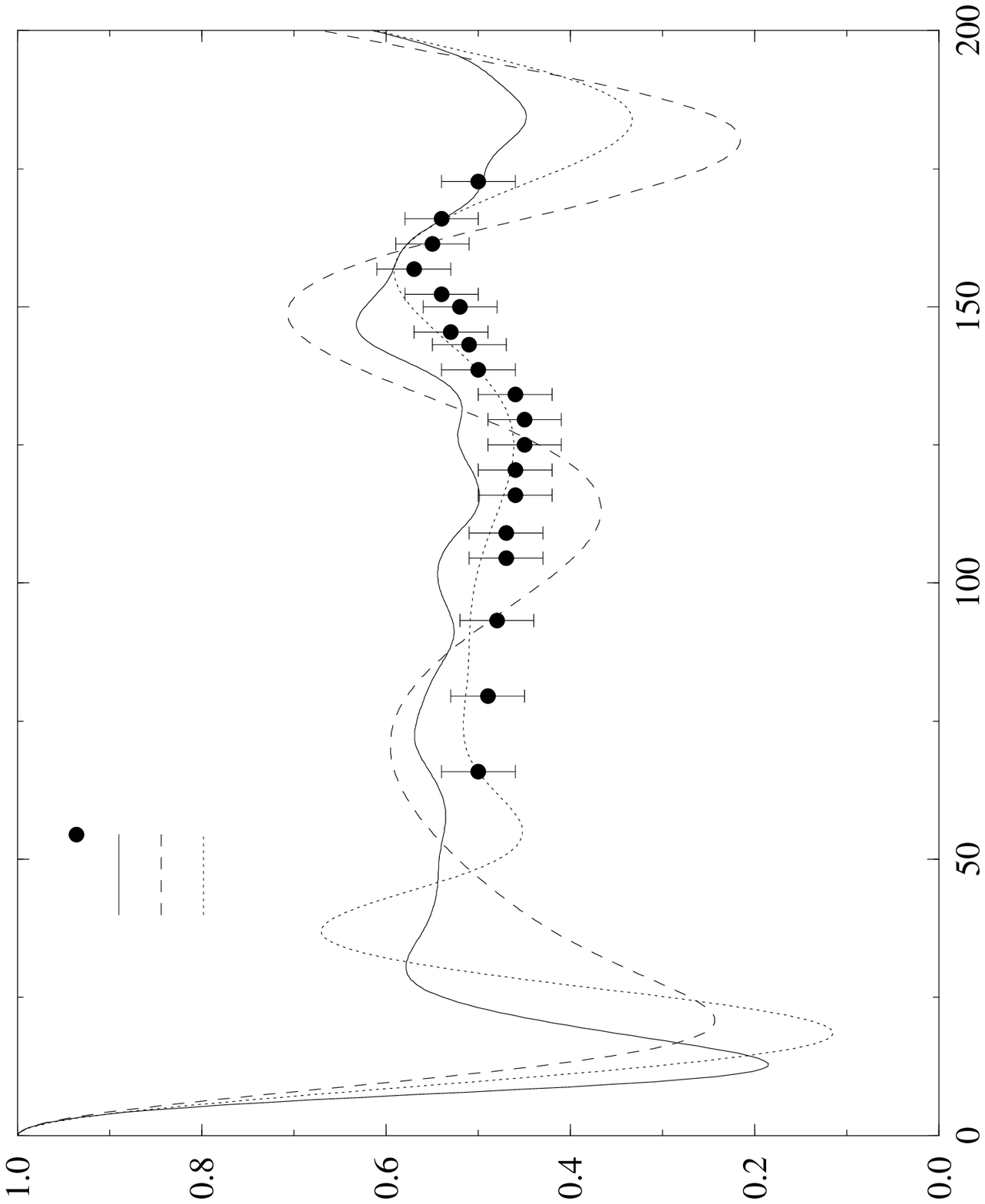}
\put(52,87){\small Experiment}
\put(52,83){\small Equilibrium: $n_b=2,~N=6$}
\put(52,79){\small Thermal: $n_b=2$}
\put(52,75){\small Poisson: $\<n\>=2.5$}
\put(70,0){\small$\tau~[\mu {\rm s}]$}
\put(0,50){\small$\P(+)$}
\put(60,23){\small$^{85}$Rb 63p$_{3/2}\leftrightarrow$ 61d$_{5/2}$}
\put(60,13){\small$R=3000~{\rm s}^{-1}$,~$n_b=2$,~$\gamma=500~{\rm s}^{-1}$}
\end{picture}
\figcap{Comparison of $\P(+)=1-\P(-)=1-\<q_{n+1}\>$ with experimental  data  of
Ref.~\cite{Rempe87}  for  various  probability  distributions.   The   Poisson
distribution   is   defined   in   Eq.~(\ref{Poisson}),    the    thermal    in
Eq.~(\ref{Thermal}),   and   the   micromaser   equilibrium   distribution   in
Eq.~(\ref{Equilibrium}). In  the  upper  figure  ($N=R/\gamma=1$)  the  thermal
distribution agrees well with the data and in the  lower  ($N=6$)  the  Poisson
distribution fits the data best. It is curious  that  the  data  systematically
seem to deviate from the micromaser equilibrium  distribution in the
lower figure.
} \label{FigDataComparison}
\end{figure}

\noi
where $M(s)$ is given  by  Eq.~(\ref{Pump})  and  $p^0$  is  the  equilibrium
distribution (\ref{Equilibrium}). The quantity $u^0$ is  a  vector  with  all
entries equal to 1, $u^0_n=1$, and represents the sum over all possible final
states of the cavity.
In Fig. \ref{FigDataComparison} we  have  compared  the
behaviour of $\P(+)$ with some characteristic experiments.

Since $\P(+)+\P(-)=1$ it is sufficient to measure the average spin value

\begin{formula}{}
\<s\>=\P(+)-\P(-)~~.
\end{formula}

\noi Since $s^2=1$ this quantity also determines the variance to be
$\<s^2\>-\<s\>^2=1-\<s\>^2$.

Correspondingly, we may define the joint probability for observing the states
of two atoms, $s_1$ followed $s_2$, with $k$ unobserved atoms between them,

\begin{formula}{kProb}
\P_k(s_1,s_2)={u^0}^\top S(s_2) S^k S(s_1) p^0~~,
\end{formula}

\noi  where  $S$  and  $S(s)$  are  defined  in  Eqs.  (\ref{Discrete1})  and
(\ref{Discrete2}).
The joint probability of finding two consecutive excited outcoming
atoms, $\P_0(+,+)$, was calculated in \cite{PaulRichter91}.
It  is  worth  noticing  that  since  $S=S(+)+S(-)$   and
$Sp^0=p^0$ we have $\sum_{s_1} \P_k(s_1,s_2)=\P(s_2)$.  Since  we  also  have
${u^0}^\top  L=  {u^0}^\top  (M-1)=0$  we  find  likewise  that   ${u^0}^\top
S={u^0}^\top$ so that  $\sum_{s_2}  \P_k(s_1,s_2)=\P(s_1)$.  Combining  these
relations we derive that $\P_k(+,-)=\P_k(-,+)$, as  expected.  Due  to  these
relations there is  essentially  only  one  two-point  function,  namely  the
``spin--spin'' covariance function

\begin{formulas}{}
\<ss\>_k&=&\sum_{s_1,s_2} s_1s_2\P_k(s_1,s_2)\\
&=&\P_k(+,+)+ \P_k(-,-)- \P_k(+,-)- \P_k(-,+)\\
&=&1-4\P_k(+,-)~~.
\end{formulas}

\noi From this we derive the properly normalized correlation function

\begin{formula}{}
\gamma^A_k=\frac{\<ss\>_k-\<s\>^2}{1-\<s\>^2}~~,
\end{formula}

\noi which satisfies $-1\le\gamma^A_k\le1$.

At large times, when $k\to\infty$, the correlation function is  in  general
expected to decay exponentially, and we define the atomic beam  correlation
length $\xi_A$ by the asymptotic behaviour for large $k\simeq Rt$

\begin{formula}{}
\gamma^A_k\sim\exp\(-\frac{k}{R\xi_A}\)~~.
\end{formula}

\noi Here we have scaled with $R$, the average  number  of  atoms  passing  the
cavity per unit of time, so that $\xi_A$ is the typical length of time that the
cavity remembers previous pumping events.


\subsection{Cavity observables}

In the context  of  the  micromaser  cavity,  one  relevant  observable  is
the
instantaneous number of photons $n$, from which we may form the average $\<n\>$
and correlations in time. The quantum state of light in  the  cavity  is  often
characterized by  the  Fano--Mandel  quality  factor  \cite{Mandel79},  which
is
related to the fluctuations of $n$ through

\begin{formula}{}
Q_f=\frac{\<n^2\>-\<n\>^2}{\<n\>}-1~~.
\end{formula}

\noi This quantity vanishes for coherent (Poisson) light  and  is  positive
for classical light.
\bort{
 (see Fig.~\ref{FigMuQf})
}

In equilibrium there is a relation between the average photon occupation number
and the spin average in the atomic beam, which is trivial to derive from the
equilibrium distribution

\begin{formula}{}
\<n\>={u^0}^\top \hat{n} p^0=n_b+N\P(-)=n_b+N\frac{1-\<s\>}{2}~~,
\end{formula}

\noi where $\hat n$ is a diagonal matrix representing the quantum number $n$. A
similar but more uncertain relation  between  the  Mandel  quality  factor  and
fluctuations in the atomic beam may also be derived \cite{Rempe90a}.

The covariance between the values of the photon occupation number $k$ atoms
apart in equilibrium is easily seen to be given by

\begin{formula}{Relation}
\<n n\>_k={u^0}^\top \hat n S^k \hat n p^0~~,
\end{formula}

\noi and again a normalized correlation function may be defined

\begin{formula}{}
\gamma^C_k=\frac{\<n n\>_k-\<n\>^2}{\<n^2\>-\<n\>^2}~~.
\end{formula}

\noi The cavity correlation length $\xi_C$ is defined by

\begin{formula}{}
\gamma^C_k\sim \exp\(-\frac{k}{R\xi_C}\)~~.
\end{formula}

\noi Since the same power of the matrix $S$  is  involved,  both  correlation
lengths are determined by  the  same  eigenvalue,  and  the  two  correlation
lengths are  therefore  identical  $\xi_A=\xi_C=\xi$  and  we  shall  no longer
distinguish between them.


\subsection{Monte Carlo determination of correlation lengths}

Since the statistical behaviour of the micromaser is a classical Markov process
it is possible to simulate it by means of Monte Carlo methods using the  cavity
occupation number $n$ as stochastic variable.

A sequence of excited atoms is generated at Poisson-distributed times and are
allowed  to  act  on  $n$   according   to   the   probabilities   given   by
Eq.~(\ref{TransEl}). In these simulations we have for simplicity chosen $a=1$
and $b=0$. After the interaction the cavity is allowed to  decay  during  the
waiting time until the next atom arrives. The action of this process  on  the
cavity variable $n$ is simulated by means  of  the  transition  probabilities
read off from the dissipative master  equation  (\ref{Dissipative})  using  a
suitably small time step $dt$. The states  of  the  atoms  in  the  beam  are
determined by the pumping transitions and the atomic correlation function may
be determined from this sequence of spin values by making  suitable  averages
after the system has reached equilibrium. Finally the correlation lengths may
be extracted numerically from the Monte Carlo data.

\begin{figure}[htb]
\unitlength=1mm
\begin{picture}(140,100)(0,0)
\includegraphics{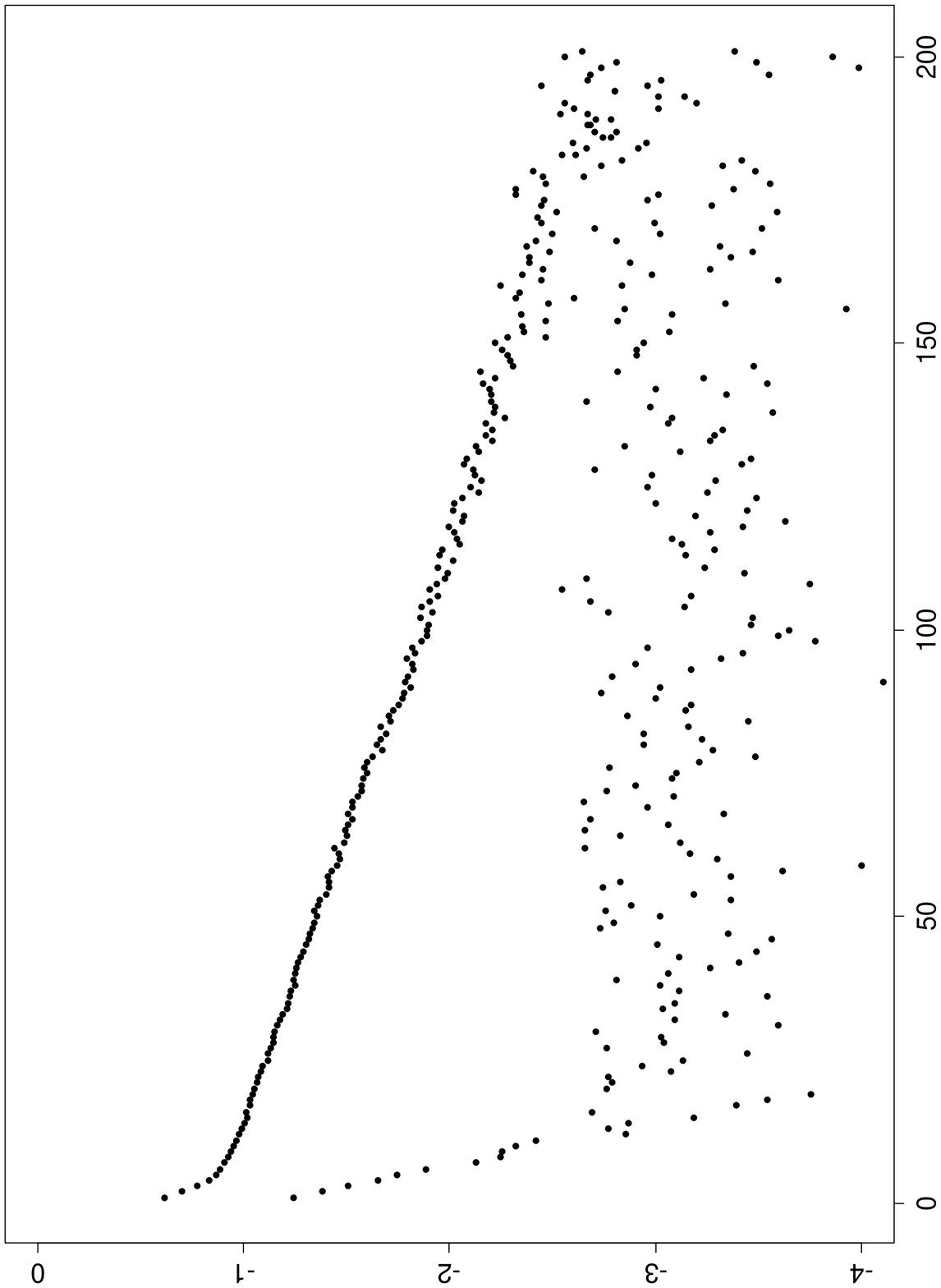}
\put(75,5){\small$k\simeq Rt$}
\put(5,53){\small$\log_{10}\gamma$}
\put(80,90){\small$^{85}$Rb 63p$_{3/2}\leftrightarrow$ 61d$_{5/2}$}
\put(80,85){\small$R=50~{\rm s}^{-1}$}
\put(80,80){$n_b=0.15$}
\put(80,75){$\gamma=5~{\rm s}^{-1}$}
\end{picture}
\figcap{Monte Carlo data (with $10^6$ simulated atoms) for the correlation as a
function of the separation $k\simeq Rt$ between the  atoms  in  the  beam  for
$\tau=25~\mu$s (lower data points) and $\tau=50~\mu$s (upper data  points).
In
the latter case the exponential  decay  at  large  times  is  clearly  visible,
whereas it is hidden in the noise in the former. The parameters  are  those  of
the experiment described in Ref. \cite{Rempe90}.
}
\label{FigFit}
\end{figure}

This extraction is, however, limited by noise due to the finite  sample  size
which in our simulation is $10^6$ atoms. In  regions  where  the  correlation
length is large,  it  is  fairly  easy  to  extract  it  by  fitting  to  the
exponential decay, whereas it is more difficult in the regions  where  it  is
small (see Fig. \ref{FigFit}). This accounts for the differences between  the
exact numerical calculations and the Monte Carlo data in Fig. \ref{FigXi}. It
is expected that real experiments will face the  same  type  of  problems  in
extracting the correlation lengths from real data.


\subsection{Numerical calculation of correlation lengths}

The micromaser equilibrium distribution is the solution of $Sp=p$,  where  $S$
is the one-atom propagation matrix (\ref{Discrete1}), so  that  $p^0$  is  an
eigenvector  of  $S$  from  the  right  with  eigenvalue  $\kappa_0=1$.   The
corresponding eigenvector  from  the  left  is  $u^0$  and  normalization  of
probabilities is expressed as  ${u^0}^\top  p^0=1$.  The  general  eigenvalue
problem concerns solutions to $Sp=\kappa p$ from the right and  $u^\top  S  =
\kappa u^\top $ from the left. It is shown below  that  the  eigenvalues  are
non-degenerate, which implies that there exists a spectral resolution  of  the
form

\begin{formula}{}
S= \sum_{\ell=0}^\infty \kappa_\ell p^\ell {u^\ell}^\top~~,
\end{formula}

\noi with eigenvalues $\kappa_\ell$ and eigenvectors  $p^\ell$  and  $u^\ell$
from right and left respectively. The long-time behaviour of the  correlation
function is governed by the next-to-leading eigenvalue  $\kappa_1<1$, and  we
see that

\begin{formula}{}
R\xi=-\frac1{\log\kappa_1}~~.
\end{formula}

The   eigenvalues   are   determined   by   the    characteristic    equation
$\det\{S-\kappa\}=0$, which may be solved  numerically.  This  procedure  is,
however, not well-defined for the  infinite-dimensional  matrix  $S$, and  in
order to evaluate the determinant we have truncated the matrix to a large and
finite-size $K\times K$ with typical $K\simeq 100$. The explicit form of $S$
in Eq.~(\ref{Discrete1}) is used, which reduces the problem to the calculation
of the determinant for a Jacobi matrix. Such a matrix  vanishes  outside  the
main diagonal and the two subleading diagonals on each side. It is  shown  in
Section~\ref{EigenvalueProblem} that the eigenvalues found from this equation
are indeed non-degenerate, real, positive and less than unity.

The next-to-leading eigenvalue is shown in Fig. \ref{FigXi} and  agrees  very
well with the Monte Carlo calculations. This figure shows a surprising amount
of structure and part of the effort in the following will  be  to  understand
this structure in detail.

It is possible to derive an exact sum rule for the reciprocal eigenvalues  (see
Appendix \ref{AppSumRule}), which yields the approximate expression:

\begin{formula}{SubSumRule}
\xi\simeq1+
\sum_{n=1}^\infty\(\frac{P_n(1-P_n)}{(1+n_b)np_n}
-\frac{1-[n_b/(1+n_b)]^n}n\)~~,
\end{formula}

\begin{figure}[htb]
\unitlength=1mm
\begin{picture}(140,100)(0,0)
\includegraphics{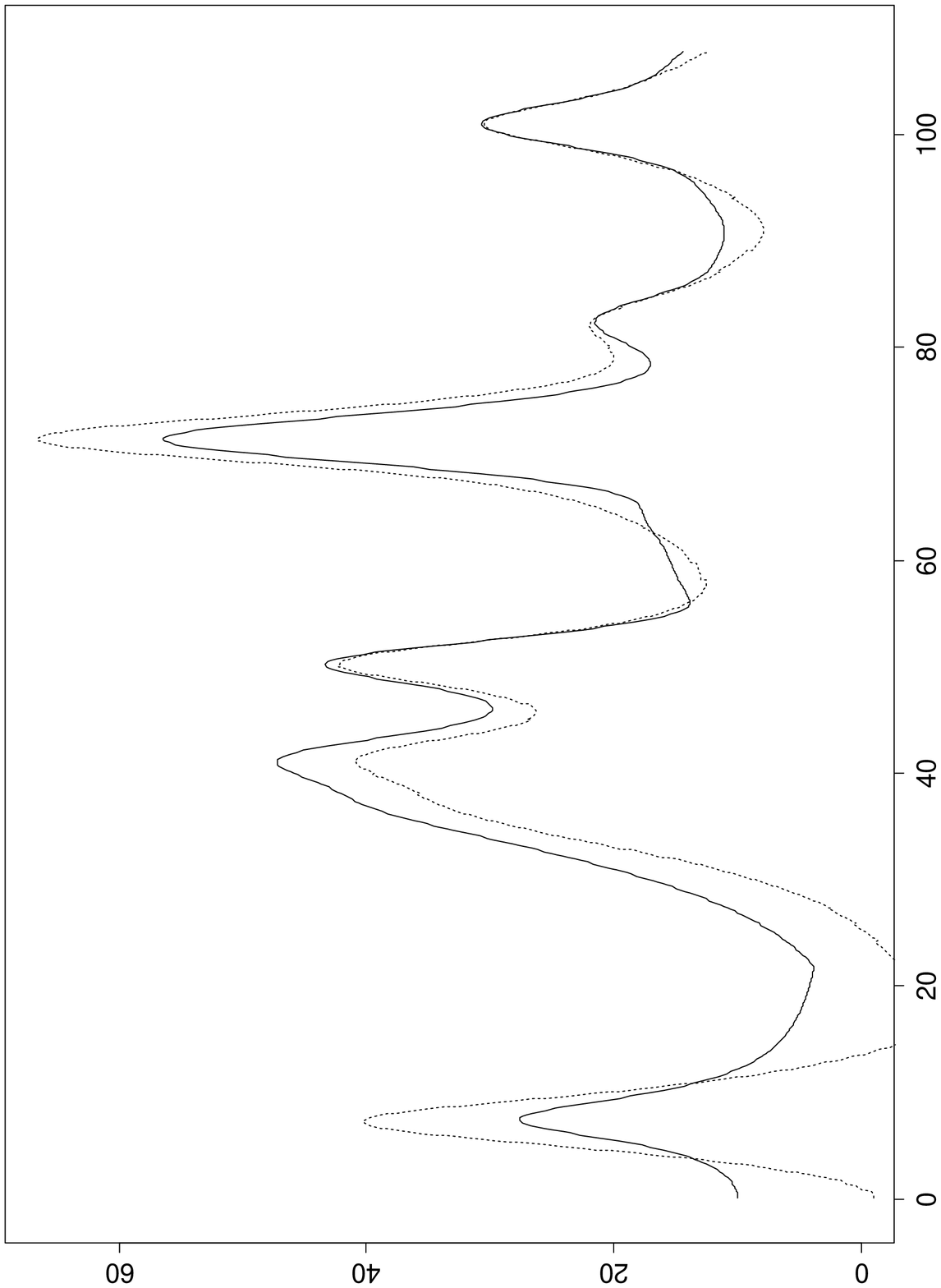}
\put(75,5){\small$\tau~[\mu {\rm s}]$}
\put(5,53){\small$R\xi$}
\put(20,90){\small$^{85}$Rb 63p$_{3/2}\leftrightarrow$ 61d$_{5/2}$}
\put(20,85){\small$R=50~{\rm s}^{-1}$}
\put(20,80){$n_b=0.15$}
\put(20,75){$\gamma=5~{\rm s}^{-1}$}
\end{picture}
\figcap{Comparison  of  the  sum  in  Eq.  (\ref{SubSumRule})  over  reciprocal
eigenvalues (dotted  curve)  with  numerically  determined  correlation  length
(solid curve) for the same parameters as in Fig.  \ref{FigXi}.  The  difference
between the curves is entirely due to the subdominant eigenvalues that have not
been taken into account here.}
\label{FigSumRule}
\end{figure}

\noi when the subdominant eigenvalues may  be  ignored.  Here  $p_n$  is  the
equilibrium distribution Eq.  (\ref{Equilibrium})  and  $P_n=\sum_{m=0}^{n-1}
p_m$ is the cumulative probability. In Fig. \ref{FigSumRule} we  compare  the
exact numerical calculation and the result of the sum rule, which is much  less
time-consuming to compute.


\section{Analytic preliminaries}
\label{analytic}
\seqnoll

In order to tackle the  task  of  determining  the  phase  structure  in  the
micromaser we need to develop  some  mathematical  tools.  The dynamics can be
formulated in two different ways which  are  equivalent  in  the  large  flux
limit. Both are related to Jacobi matrices describing the stochastic process.
Many characteristic features of the correlation length are related to scaling
properties for $N\goto\infty$,  and  require  a  detailed  analysis  of  the
continuum limit. Here we introduce some of the concepts that are used in  the
main analysis in Section \ref{Phasestructure}.

\subsection{Continuous master equation}

When the atoms have Poisson distributed  arrival  times  it  is  possible  to
formulate the problem as a differential equation \cite{Lugiato87}. Each  atom
has the same probability $Rdt$ of arriving in an infinitesimal time  interval
$dt$. Provided the interaction with the cavity takes less time  than  this
interval,  \ie\  $\tau\ll  dt$,  we  may  consider  the  transition   to   be
instantaneous and write the transition matrix as $Rdt(M-1)$ so that we get

\begin{formula}{}
\frac{dp}{dt}=-\gamma L_Cp+R(M-1)p\equiv-\gamma Lp~~,
\end{formula}

\noi where $L=L_C-N(M-1)$. This equation obviously has the solution

\begin{formula}{}
p(t)=e^{-\gamma L t}p~~.
\end{formula}

\noi Explicitly we have

\begin{formulas}{}
L_{nm}&=&(n_b+1)(n\delta_{n,m}-(n+1)\delta_{n+1,m})
+n_b((n+1)\delta_{n,m}-n\delta_{n,m+1}) \\
&&+N(
(aq_{n+1}+bq_n)\delta_{n,m}
-aq_n\delta_{n,m+1}
-bq_{n+1}\delta_{n+1,m})
{}~~,
\end{formulas}

\noi and

\begin{formulas}{Maser}
{\displaystyle\frac1\gamma\frac{dp_n}{dt}}
&=&-(n_b+1)(np_n-(n+1)p_{n+1})-n_b((n+1)p_n-np_{n-1})\\
&&-N((aq_{n+1}+bq_n)p_n-aq_np_{n-1}-bq_{n+1}p_{n+1})~~.
\end{formulas}

The equilibrium distribution may be found by the same technique as before,
writing the right-hand side of \eq{Maser} as $J_{n+1}-J_n$ with

\begin{formula}{}
J_n=((n_b+1)n+Nbq_n)p_n-(n_bn+Naq_n)p_{n-1}~~,
\end{formula}

\noi and setting $J_n=0$ for all $n$. The equilibrium distribution is  clearly
given by the same expression (\ref{Equilibrium}) as in the discrete case.


\subsection{Relation to the discrete case}

Even if the discrete and continuous  formulation  has  the  same  equilibrium
distribution, there is a difference in the dynamical  behaviour  of  the  two
cases. In the discrete case the  basic  propagation  matrix  is  $S^k$, where
$S=(1+L_C/N)^{-1}M$, whereas it is $\exp(-\gamma Lt)$ in the continuous  case.
For high pumping rate $N$ we expect the two formalisms to coincide,  when  we
identify $k\simeq Rt$.  For  the  long-time  behaviour  of  the  correlation
functions this implies that the next-to-leading eigenvalues $\kappa_1$ of $S$
and $\lambda_1$ of  $L$  must  be  related  by  $1/\xi=\gamma\lambda_1\simeq
-R\log\kappa_1$.

To prove this, let us compare the two eigenvalue problems. For the continuous
case we have

\begin{formula}{}
(L_C-N(M-1))p=\lambda p~~,
\end{formula}

\noi whereas in the discrete case we may rewrite $Sp=\kappa p$ to become

\begin{formula}{}
\(L_C -\frac{N}{\kappa}(M-1)\)p=N\(\frac1\kappa-1\)p~~.
\end{formula}

\noi Let a  solution  to  the  continuous  case  be  $p(N)$  with  eigenvalue
$\lambda(N)$, making explicit the dependence on $N$. It is then obvious  that
$p(N/\kappa)$ is a solution to the discrete case with eigenvalue $\kappa$
determined by

\begin{formula}{EigenRelation}
\lambda\(\frac N\kappa\)=N\(\frac1\kappa-1\)~~.
\end{formula}

\noi As we shall  see  below,  for  $N\gg1$  the  next-to-leading  eigenvalue
$\lambda_1$ stays finite or goes to zero, and hence $\kappa_1\to1$  at  least
as fast as $1/N$. Using this result it follows that the correlation length is
the same to $\O(1/N)$ in the two formalisms.


\subsection{The eigenvalue problem}\label{EigenvalueProblem}

The transition matrix $L$ truncated to size  $(K+1)\times(K+1)$  is  a  special
kind of asymmetric Jacobi matrix

\begin{formula}{LK}
L_K=\left\{\begin{array}{ccccccc}
A_0+B_0&-B_1&0&0&\cdots\\
-A_0&A_1+B_1&-B_2&0&\cdots\\
0&-A_1&A_2+B_2&-B_3\\
\vdots&\vdots&\vdots&\vdots&\vdots\\
&&&&-A_{K-2}&A_{K-1}+B_{K-1}&-B_{K}\\
&&&\cdots&0&-A_{K-1}&A_K+B_K\\
\end{array}\right\}~,
\end{formula}

\noi where

\begin{formulas}{}
A_n&=&n_b(n+1)+Naq_{n+1}~~,\\
B_n&=&(n_b+1)n+Nbq_n~~.
\end{formulas}

\noi Notice that the sum over the elements in every column  vanishes,  except
for the first and the last, for which the sums respectively take  the  values
$B_0$ and $A_K$. In our case we have $B_0=0$,  but  $A_K$  is  non-zero.  For
$B_0=0$ it is easy to see  (using  row  manipulation)  that  the  determinant
becomes  $A_0A_1\cdots  A_K$  and  obviously  diverges  in   the   limit   of
$K\to\infty$.  Hence  the  truncation  is  absolutely  necessary.   All   the
coefficients in the characteristic equation diverge, if we do not  truncate.
In order to secure that there is an eigenvalue $\lambda=0$,  we  shall  force
$A_K=0$ instead of the value given above. This means that the matrix  is  not
just truncated but actually changed in the last diagonal element.  Physically
this secures that
there is no external input to the process from cavity occupation
numbers above $K$, a not unreasonable requirement.

An eigenvector to the right satisfies the  equation  $L_Kp=\lambda p$,  which
takes the explicit form

\begin{formula}{}
-A_{n-1}p_{n-1}+(A_n+B_n)p_n -B_{n+1}p_{n+1}=\lambda p_n~~.
\end{formula}

\noi Since we may solve this equation successively for $p_1,  p_2,\ldots,  p_K$
given  $p_0$,  it  follows  that  all  eigenvectors  are  non-degenerate.   The
characteristic polynomial obeys the recursive  equation

\begin{formula}{}
    \det(L_K-\lambda)=(A_K+B_K-\lambda)
    \det(L_{K-1}-\lambda)-A_{K-1}B_K\det(L_{K-2}-\lambda)~~,
\end{formula}

\noi and this is also the characteristic  equation  for  a  symmetric  Jacobi
matrix  with  off-diagonal  elements  $C_n=-\sqrt{A_{n-1}B_n}$.   Hence   the
eigenvalues are the same and therefore all real and, as we  shall  see  below,
non-negative. They may therefore be ordered $0= \lambda_0 < \lambda_1 <\cdots
<\lambda_K$. The equilibrium distribution (\ref{Equilibrium}) corresponds  to
$\lambda=0$ and is given by

\begin{formula}{EquilAB}
p^0_n=p^0_0\prod_{m=1}^n \frac{A_{m-1}}{B_m}=p^0_0
\frac{A_0A_1\cdots A_{n-1}}{B_1B_2\cdots B_n}
{}~~~~\mbox{for $n=1,2,\ldots,K$}~~.
\end{formula}

\noi Notice that this  expression  does  not  involve  the  vanishing  values
$B_0=A_K=0$.

Corresponding to each eigenvector $p$ to the right there  is  an  eigenvector
$u$ to the left, satisfying $u^\top L_K=\lambda u^\top$, which in  components
reads

\begin{formula}{}
A_n(u_n-u_{n-1})+B_n(u_n-u_{n+1})=\lambda u_n~~.
\end{formula}

\noi For $\lambda=0$ we obviously have $u^0_n=1$ for all $n$ and  the  scalar
product $u^0\cdot p^0=1$. The eigenvector to the left is trivially related to
the eigenvector to the right via the equilibrium distribution

\begin{formula}{uRelation}
p_n=p^0_nu_n~~.
\end{formula}

\noi  The  full  set  of  eigenvectors  to  the   left   and   to   the   right
$\{u^\ell,p^\ell~|~\ell=0,1,2,\ldots,K\}$ may now be chosen to  be  orthonormal
$u^\ell\cdot p^{\ell'}=\delta_{\ell,\ell'}$, and is, of course, complete  since
the dimension $K$ is finite.

It is useful to  express  this  formalism  in  terms  of  averages  over  the
equilibrium  distribution  $\<f_n\>_0=\sum_{n=0}^K  f_n  p^0_n$.  Then  using
Eq.~(\ref{uRelation})  we  have,  for  an  eigenvector  with  $\lambda>0$, the
relations

\begin{formulas}{}
\<u_n\>_0&=&0~~,\\
\<u_n^2\>_0&=&1~~,\\
\<u_nu'_n\>_0&=&0&\mbox{for $\lambda\ne\lambda'$}~~.
\end{formulas}

\noi Thus the eigenvectors with $\lambda>0$ may  be  viewed  as  uncorrelated
stochastic functions of $n$ with zero mean and unit variance.

Finally, we rewrite the eigenvalue equation to  the  right  in  the  form  of
$\lambda p_n=J_n-J_{n+1}$ with

\begin{formula}{}
J_n=B_np_n-A_{n-1}p_{n-1}=p^0_nB_n(u_n-u_{n-1})~~.
\end{formula}

\noi Using the orthogonality we then find

\begin{formula}{}
\lambda=\sum_{n=0}^K u_n (J_n-J_{n+1})=\<B_n(u_n-u_{n-1})^2\>_0~~,
\end{formula}

\noi which incidentally proves that all eigenvalues are non-negative.  It  is
also evident that an eigenvalue is built up from the non-constant parts, \ie\
the jumps of $u_n$.

\subsection{Effective potential}

It is convenient to introduce an effective potential  $V_n$,
first discussed by Filipowicz {\it et al.} \cite{Filipowicz86} in the
continuum limit,   by  writing  the
equilibrium distribution (\ref{Equilibrium}) in the form

\begin{formula}{}
p_n=\frac1Ze^{-NV_n}~~,
\end{formula}

\noi with

\begin{formula}{ExactPot}
V_n=-\frac1N\sum_{m=1}^n\log\frac{n_bm+Naq_m}{(1+n_b)m+Nbq_m}~~,
\end{formula}

\noi for $n\ge1$. The  value  of  the  potential  for  $n=0$  may  be  chosen
arbitrarily, for example $V_0=0$, because of the normalization constant

\begin{formula}{}
Z=\sum_{n=0}^\infty e^{-NV_n}~~.
\end{formula}

\noi It is, of course, completely  equivalent  to  discuss  the  shape  of  the
equilibrium distribution and the shape of the effective potential.
Our definition of $V_n$ differs from the one introduced in
Refs.~\cite{Filipowicz86,Guzman89}
in the sense that our $V_n$ is exact while the one in
\cite{Filipowicz86,Guzman89}
was derived from a Fokker-Planck equation in the continuum limit.


\subsection{Semicontinuous formulation}

Another way of making analytical methods, such  as
the Fokker--Planck equation,
easier to use is to rewrite the formalism (exactly) in  terms  of
the  scaled  photon  number  variable $x$ and the scaled time parameter
$\theta$, defined by
\cite{Filipowicz86}

\begin{formulas}{Defs}
x&=&\displaystyle{\frac nN}~~,\\
\theta&=&g\tau\sqrt N~~.\\
\end{formulas}

\noi Notice that the variable $x$ and not $n$  is  the  natural  variable  when
observing  the  field  in  the  cavity  by  means  of  the  atomic  beam   (see
(\ref{Relation})).  Defining  $\Delta  x=1/N$  and   introducing   the   scaled
probability distribution $p(x)=Np_n$ the conservation of probability takes  the
form

\begin{formula}{}
\sum_{x=0}^\infty \Delta x\; p(x)=1~~,
\end{formula}

\noi where the sum extends over all discrete values of $x$ in the interval.
Similarly the equilibrium distribution  takes the form

\begin{formula}{}
p^0(x)=\frac1Z_x e^{-NV(x)}~~,
\end{formula}

\noi with the effective potential given as an ``integral''

\begin{formula}{Effective}
V(x)=\sum_{x'>0}^x \Delta x'\;D(x')~~,
\end{formula}

\noi with ``integrand''

\begin{formula}{Derivative}
D(x)=-\log\frac{n_bx+aq(x)}{(1+n_b)x+bq(x)}~~.
\end{formula}

\noi The transition probability function is $q(x)=\sin^2\theta\sqrt x$
and the normalization constant is given by

\begin{formula}{}
Z_x=\frac ZN=\sum_{x=0}^\infty\Delta x\;e^{-NV(x)}~~.
\end{formula}

In order to reformulate the master equation (\ref{Maser}) it is convenient
to introduce the discrete derivatives $\Delta_+f(x)=f(x+\Delta x)-f(x)$ and
$\Delta_-f(x)=f(x)-f(x-\Delta x)$. Then we find

\begin{formula}{}
\frac1\gamma\frac{dp(x)}{dt}=\frac{\Delta_+}{\Delta x}J(x)~~,
\end{formula}

\noi with

\begin{formula}{}
J(x)=(x-(a-b)q(x))p(x)+\frac1N(n_bx+aq(x))\frac{\Delta_-}{\Delta x}p(x)~~.
\end{formula}

\noi
For  the  general  eigenvector  we  define  $p(x)=Np_n$   and   write   it   as
$p(x)=p^0(x)u(x)$ with $u(x)=u_n$ and find the equations

\begin{formula}{pEigen}
\lambda p(x)=-\frac{\Delta_+}{\Delta x}J(x)~~,
\end{formula}

\noi and

\begin{formula}{uEq}
J(x)=\frac1N p^0(x) ((1+n_b)x+bq(x))\frac{\Delta_-}{\Delta x}u(x)~~.
\end{formula}

\noi Equivalently the eigenvalue equation for $u(x)$ becomes

\begin{formula}{uEigen1}
\lambda u(x)=(x-(a-b)q(x))\frac{\Delta_-}{\Delta x}u(x)
-\frac1N\frac{\Delta_+}{\Delta x}
\left[(n_bx+aq(x))\frac{\Delta_-}{\Delta x}u(x)\right]~~.
\end{formula}

\noi As before we also have

\begin{formulas}{}
\<u(x)\>_0&=&0~~,\\
\<u(x)^2\>_0&=&1~~,
\end{formulas}

\noi where now the average over $p^0(x)$ is defined as
$\<f(x)\>_0=\sum_x \Delta x f(x) p^0(x)$.
As before we may also express the eigenvalue as an average

\begin{formula}{Eigen1}
\lambda=\frac1N
\left\<((1+n_b)x+bq(x))\(\frac{\Delta_-u(x)}{\Delta x}\)^2\right\>_0~~.
\end{formula}

\noi
Again it should be emphasized that all these formulas are exact rewritings  of
the previous ones,  but  this  formulation  permits  easy  transition  to  the
continuum case, wherever applicable.


\subsection{Extrema of the continuous potential}
\label{s:Extrema}

The quantity $D(x)$ in Eq.~(\ref{Derivative}) has a natural continuation to all
real values of $x$ as a  smooth  differentiable  function.  The  condition  for
smoothness is that the change in  the  argument  $\theta\sqrt  x$  between  two
neighbouring values, $x$ and $x+\Delta x$ is much  smaller than 1,  or
$\theta\ll
2N\sqrt x$. Hence for $N\to\infty$ the function is smooth everywhere and the
sum in Eq.~(\ref{Effective}) may be replaced by an integral

\begin{formula}{ContPot}
V(x)=\int_0^x dx'\; D(x')~~,
\end{formula}

\begin{figure}[p]
\unitlength=1mm

\begin{picture}(140,90)(0,0)
\includegraphics{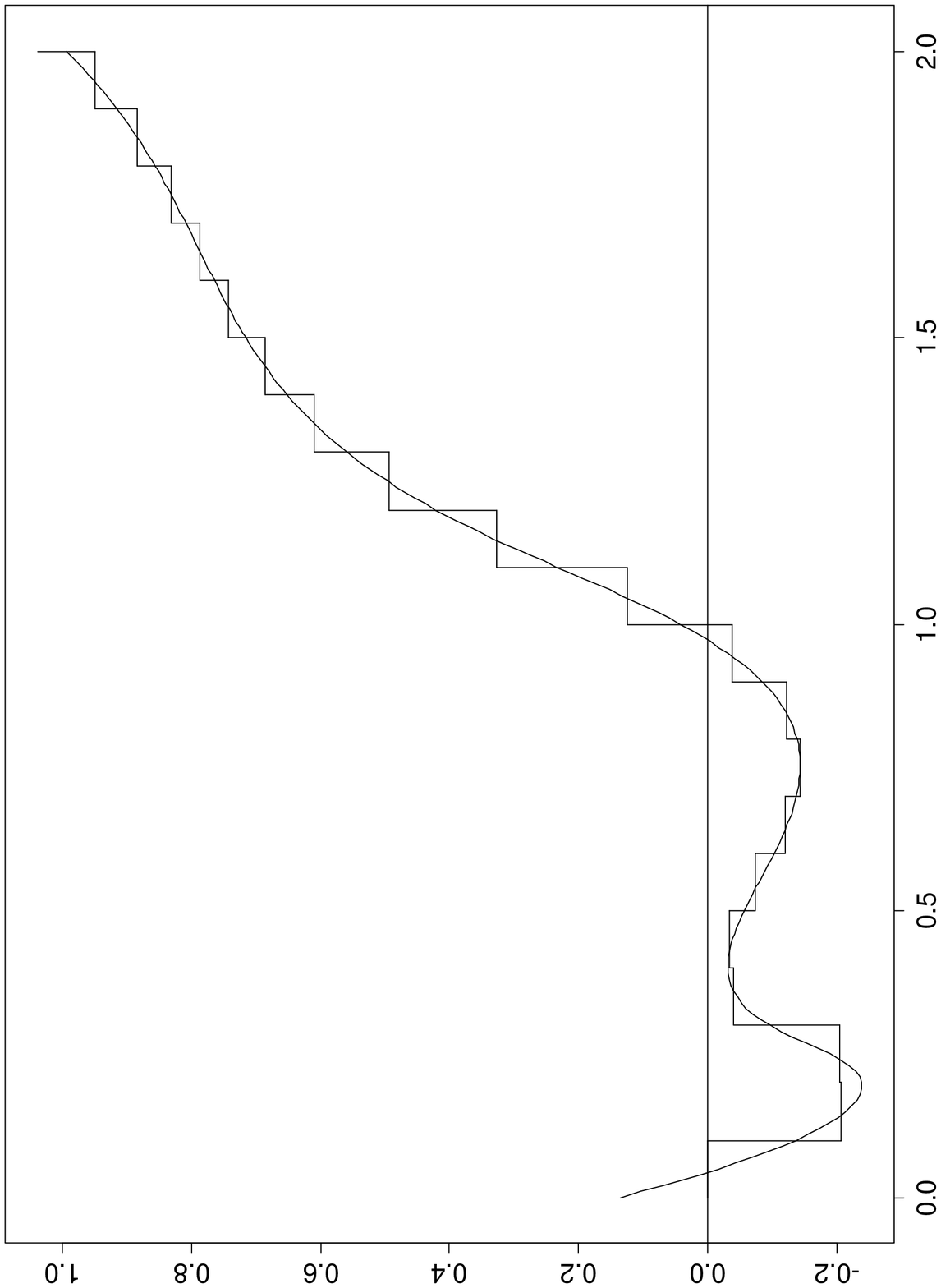}
\put(30,80){\small\parbox{30mm}{$\theta=6.0$\\$N=10$\\$n_b=0.15$\\$a=1,~b=0$}}
\put(34,20){\small$x_0$}
\put(45,32){\small$x_1$}
\put(64,24){\small$x_2$}
\put(0,50){\small$V(x)$}
\end{picture}

\begin{picture}(140,90)(0,0)
\includegraphics{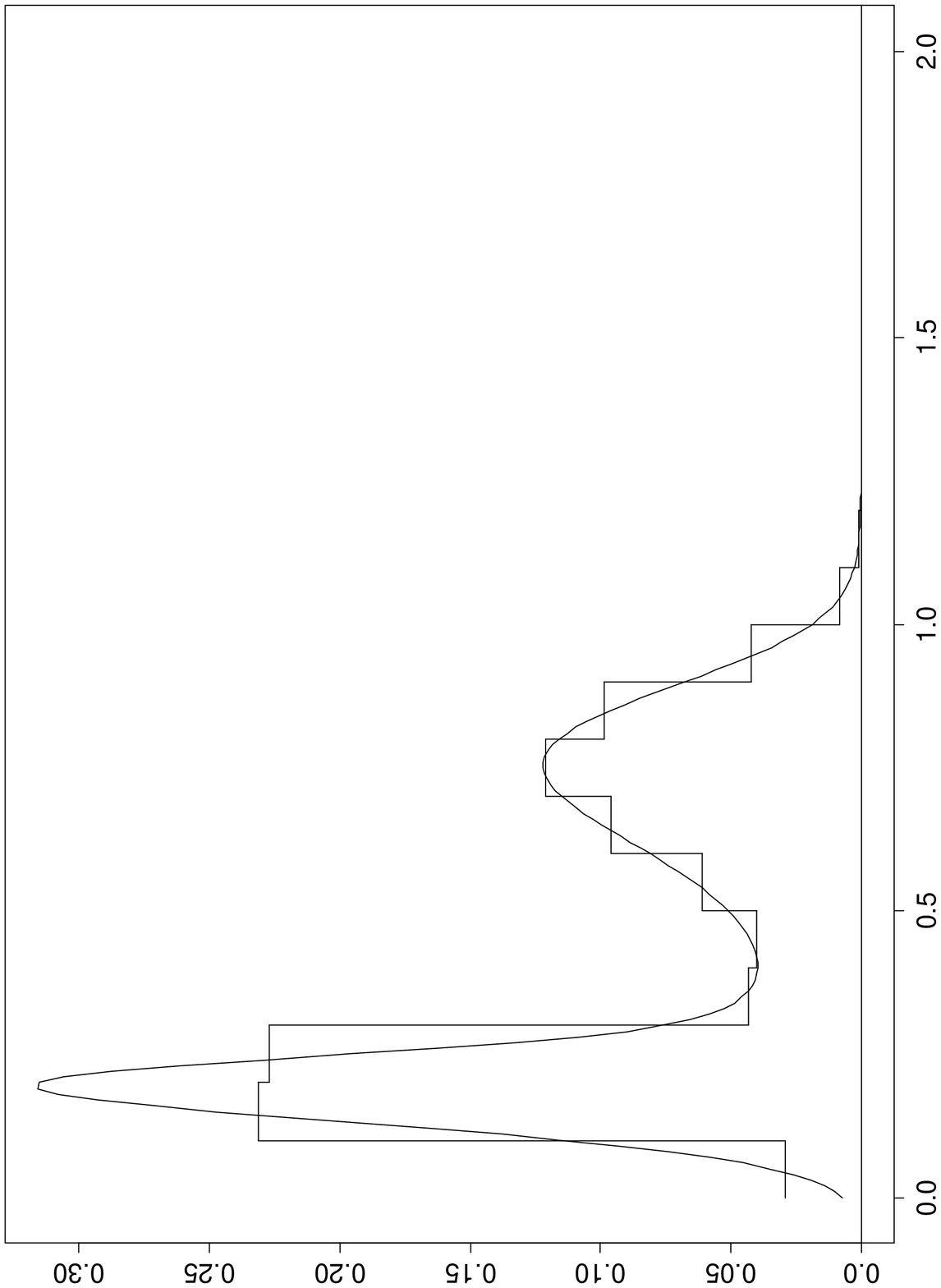}
\put(100,80){\small\parbox{30mm}{$\theta=6.0$\\$N=10$\\$n_b=0.15$\\$a=1,~b=0$}}
\put(0,50){\small$p(x)$}
\put(78,3){\small$x$}
\end{picture}

\figcap{Example of a potential with two minima $x_0,x_2$ and one maximum $x_1$
(upper  graph).  The  rectangular  curve   represents   the   exact   potential
(\ref{ExactPot}),
whereas the continuous curve is given by Eq. (\ref{Effective})
with the summation replaced  by  an  integral.  The  value  of  the  continuous
potential at $x=0$ has been chosen such as to make the distance minimal between
the two curves. In the lower graph the corresponding  probability  distribution
is shown. }
\label{FigEffPot}
\end{figure}

\noi so that  $D(x)=V'(x)$. In Fig. \ref{FigEffPot} we illustrate the typical
behaviour of the potential and the corresponding photon number distribution
in the first critical region (see Section \ref{FirstCritical}). Notice that
the photon-number distribution exhibits Schleich--Wheeler oscillations typical
of a squeezed state \cite{Schleich87}.

 The extrema of this potential are located at
the solutions to $q(x)=x$; they may be parametrized in the form

\begin{formulas}{Extrema}
x&=&(a-b)\sin^2\phi~~,\\
\theta&=&\displaystyle{\frac1{\sqrt{a-b}}~\frac\phi{|\sin\phi|}}~~,\\
\end{formulas}

\noi with $0\le\phi<\infty$. These formulas map out a multibranched function
$x(\theta)$ with critical points where the derivative

\begin{formula}{}
D'(x)=V''(x)=\frac{(a+n_b(a-b))(q(x)-xq'(x))}
{((1+n_b)x+bq(x))(n_bx+aq(x))}
\end{formula}

\noi  vanishes,  which   happens   at   the   values   of   $\phi$   satisfying
$\phi=\tan\phi$.   This   equation    has    an    infinity    of    solutions,
$\phi=\phi_k,~~k=0,1,\ldots$, with $\phi_0=0$ and to a good approximation

\begin{formula}{}
\phi_k=(2k+1)\frac\pi2-\frac1{(2k+1)\frac\pi2}
+\O\(\((2k+1)\frac\pi2\)^{-3}\)
\end{formula}

\noi for $k=1,2,\ldots$, and each of these branches  is  double-valued,
with a sub-branch
corresponding to a minimum ($D'>0$) and another  corresponding
to a maximum ($D'<0$). Since there are always $k+1$ minima and  $k$  maxima, we
denote the minima $x_{2k}(\theta)$ and the maxima $x_{2k+1}(\theta)$. Thus  the
minima have even indices and the maxima have odd indices. They are given  as  a
function of $\theta$ through Eq.~(\ref{Extrema}) when $\phi$ runs  through
certain
intervals. Thus, for the minima of $V(x)$, we have

\begin{formula}{}
\phi_k<\phi<(k+1)\pi,~~\theta_k<\theta<\infty,~~a-b>x_{2k}(\theta)>0,
{}~~~k=0,1,\ldots~~,
\end{formula}

\noi
and for the maxima

\begin{formula}{}
k\pi<\phi<\phi_k,~~\infty>\theta>\theta_k,~0<x_{2k+1}(\theta)<a-b,~~k=1,\ldots
\end{formula}

\noi Here $\theta_k=\phi_k/|\sin\phi_k|\sqrt{a-b}$ is the value of $\theta$ for
which the
$k$'th   branch   comes    into    existence.    Hence    in    the    interval
$\theta_K<\theta<\theta_{K+1}$   there    are    exactly    $2K+1$    branches,
$x_0,x_1,x_2,\ldots,x_{2K-1},x_{2K}$, forming the $K+1$ minima and  $K$  maxima
of $V(x)$. For $0<\theta<\theta_0=1/\sqrt{a-b}$ there are no extrema.

This classification allows us to discuss the different  parameter  regimes
that arise in the limit of $N\to\infty$. Each  regime  is  separated  from  the
others by singularities and are thus equivalent to the phases that arise in the
thermodynamic limit of statistical mechanics.



\section{Phase structure}
\seqnoll
\label{Phasestructure}

We shall from now on limit the discussion to the case of  initially  completely
excited  atoms,  $a=1,~b=0$,  which   simplifies   the   following   discussion
considerably.

\begin{figure}[htb]
\unitlength=1mm
\begin{picture}(140,100)(0,0)
\includegraphics{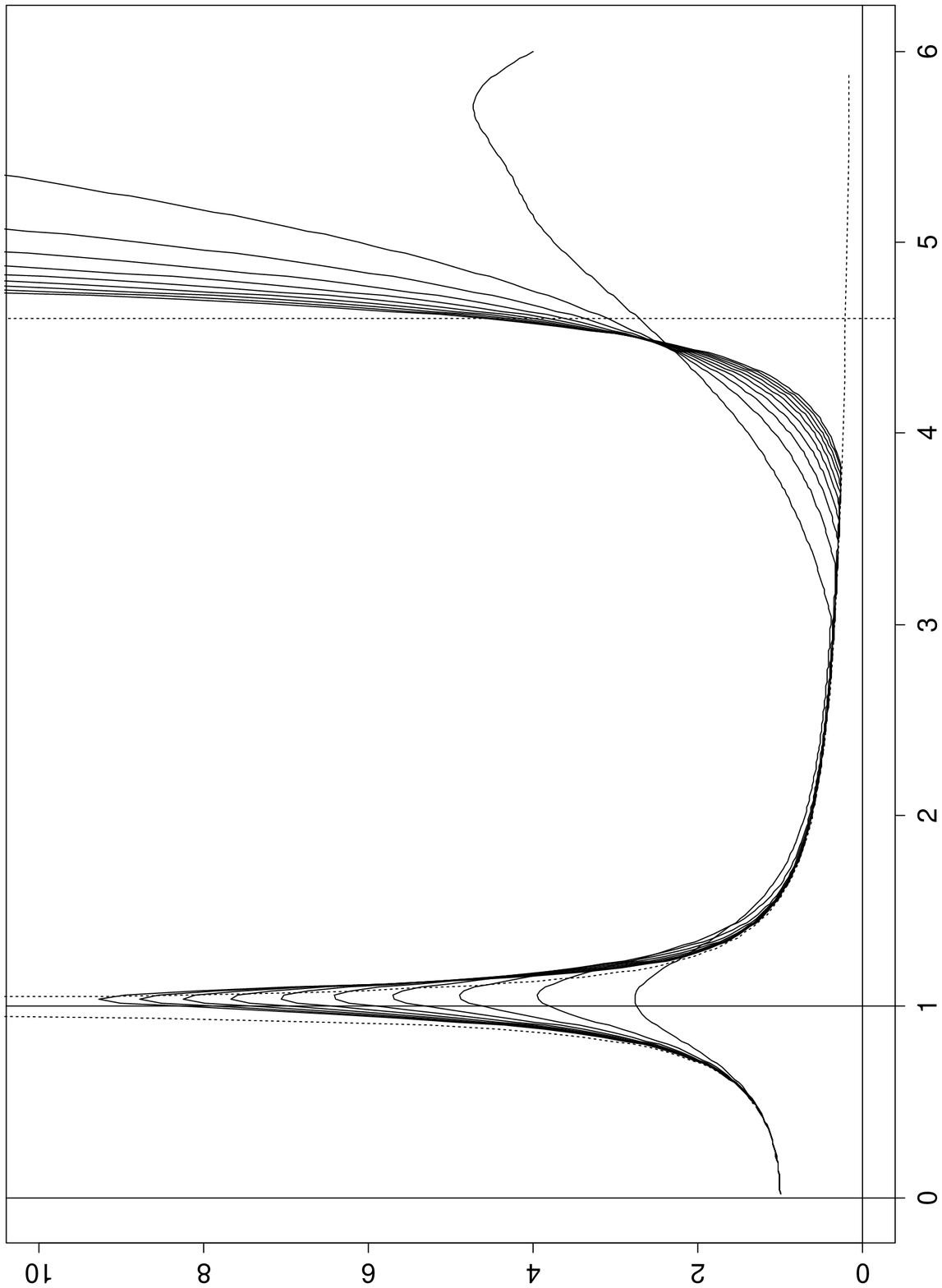}
\put(75,5){\small$\theta$}
\put(5,53){\small$\gamma\xi$}
\put(50,85){\small$N=10,20,\ldots,100$}
\put(50,80){\small$n_b=0.15$}
\end{picture}
\figcap{The correlation  length  in  the  thermal  and  maser
phases as a function of $\theta$ for various values of $N$. The  dotted  curves
are the limiting value for $N=\infty$. The correlation length grows  as  $\sqrt
N$ near $\theta=1$ and exponentially for $\theta>\theta_1\simeq4.603$. }
\label{FigMaser}
\end{figure}

\begin{figure}[htb]
\unitlength=1mm
\begin{picture}(140,90)(0,0)
\includegraphics{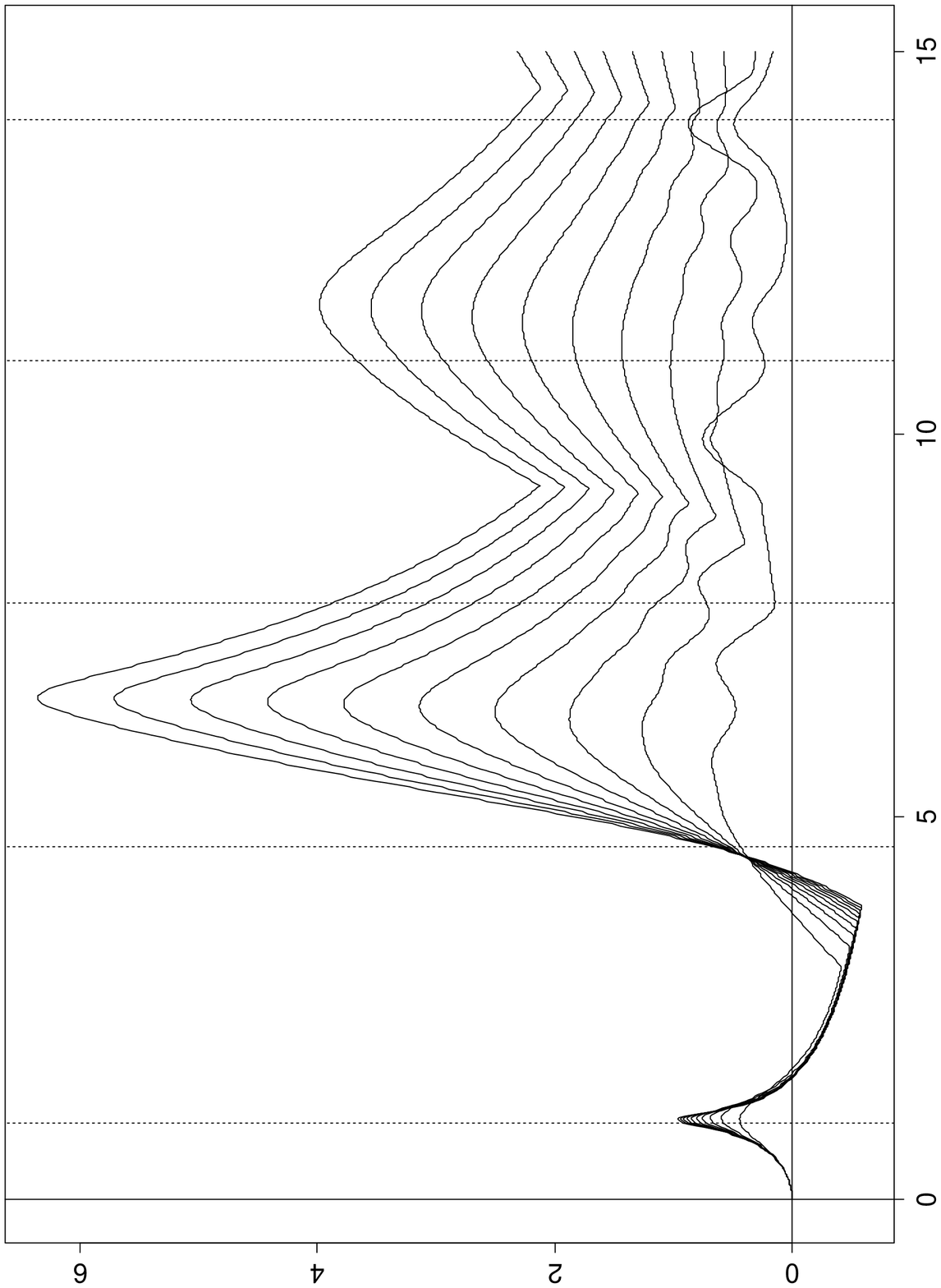}
\put(75,0){\small$\theta$}
\put(5,53){\small$\log(\gamma\xi)$}
\end{picture}
\figcap{The logarithm of the correlation length as a function of  $\theta$  for
various values of $N$
$(10,20,\ldots,100)$. We have $n_b=0.15$  here.  Notice  that
for $\theta>\theta_1$ the logarithm of the correlation  length  grows  linearly
with $N$ for large $N$.  The vertical lines indicate
$\theta_0=1,~\theta_1=4.603,
{}~ \theta_2=7.790, ~\theta_3=10.95$ and $\theta_4=14.10$.
}
\label{FigCritic}
\end{figure}

The central issue in this paper is  the  phase  structure  of  the  correlation
length as a function of the parameter $\theta$. In the limit of infinite atomic
pumping rate, $N\to\infty$, the statistical  system  described  by  the  master
equation (\ref{Master}) has a number of different dynamical  phases,  separated
from each other by singular boundaries in the space of parameters. We shall  in
this section investigate the character of the different  phases,  with  special
emphasis on the limiting behaviour of the correlation length. There  turns  out
to be several qualitatively different phases within a range of  $\theta$  close
to experimental values. First, the thermal phase  and  the  transition  to  the
maser phase at $\theta=1$ has previously been discussed  in  terms  of  $\<n\>$
\cite{Filipowicz86,Lugiato87,Guzman89}. The  new  transition  to  the  critical
phase  at  $\theta_1\simeq4.603$  is  not  revealed  by  $\<n\>$  and  the
introduction of the correlation length as an observable is necessary to
describe
it. In the large flux limit $\<n\>$ and $\<(\Delta n)^2\>$ are  only  sensitive
to the probability distribution close to its global  maximum.  The  correlation
length depends crucially also on local  maxima  and  the  phase  transition  at
$\theta_1$ occurs when a new local maximum emerges. At $\theta\simeq 6.3$ there
is a phase transition in $\<n\>$ taking a discrete jump to a higher  value.  It
happens when there are two competing global minima in the  effective  potential
for different values of $n$. At the same point the correlation  length  reaches
its maximum. In Fig. \ref{FigMaser} we  show  the  correlation  length  in  the
thermal and maser phases, and in Fig. \ref{FigCritic} the critical phases,  for
various values of the pumping rate $N$.

\subsection{Empty cavity}

When there is no interaction, \ie\ $M=1$, or equivalently $q_n=0$ for all  $n$,
the behaviour of the cavity is purely thermal, and then it is possible to  find
the eigenvalues explicitly. Let us in this case write

\begin{formula}{}
L_C=(2n_b+1)L_3-(1+n_b)L_--n_bL_+-\frac12~~,
\end{formula}

\noi
where

\begin{formulas}{}
(L_3)_{nm}&=&\(n+\frac12\)\delta_{nm}~~,\\
(L_+)_{nm}&=&n\delta_{n,m+1}~~,\\
(L_-)_{nm}&=&(n+1)\delta_{n+1,m}~~.\\
\end{formulas}

\noi
These operators form a representation of the Lie algebra of SU(1,1)

\begin{formula}{}
[L_-,L_+]=2L_3~, \quad [L_3,L_\pm]=\pm L_\pm~~.
\end{formula}

\noi It then follows that

\begin{formula}{}
L_C=e^{rL_+}e^{-(1+n_b)L_-}(L_3-\textstyle{\frac12})\,e^{(1+n_b)L_-}e^{-rL+}~~,
\end{formula}

\noi
where $r=n_b/(1+n_b)$. This proves that $L_C$ has the same eigenvalue  spectrum
as the simple number operator $L_3-\textstyle{\frac12}$, \ie\ $\lambda_n=n$ for
$n=0,1,\ldots$. Since $M=1$ for $\tau=0$  this  is  a  limiting  case  for  the
correlation lengths  $\gamma\xi_n=1/\lambda_n=1/n$  for  $\theta=0$.  From  Eq.
(\ref{EigenRelation}) we obtain  $\kappa_n=1/(1+n/N)$  in  the  non-interacting
case. Hence  in  the  discrete  case  $R\xi_n=-1/\log\kappa_n\simeq  N/n$  for
$N\gg n$ and this agrees with the values in Fig. \ref{FigXi} for $n=1,2,3$
near
$\tau=0$.


\subsection{Thermal phase: $0\le\theta<1$}

In this phase  the  natural  variable  is  $n$,  not  $x=n/N$.  The  effective
potential has no extremum for $0<n<\infty$, but is smallest  for  $n=0$.  Hence
for $N\to\infty$ it may be approximated by its leading linear  term  everywhere
in this region

\begin{formula}{}
NV_n=n\log\frac{n_b+1}{n_b+\theta^2}~~.
\end{formula}

\noi Notice that the slope vanishes for $\theta=1$. The higher-order terms play
no  role  as  long  as  $1-\theta^2\gg1/\sqrt{N}$,  and  we  obtain  a   Planck
distribution

\begin{formula}{}
p^0_n=\frac{1-\theta^2}{1+n_b}\(\frac{n_b+\theta^2}{1+n_b}\)^n~~,
\end{formula}

\noi  with photon number average

\begin{formula}{}
\<n\>=\frac{n_b+\theta^2}{1-\theta^2}~~,
\end{formula}

\noi which (for $\theta>0$) corresponds to an increased temperature.  Thus  the
result of pumping the cavity with the  atomic  beam  is  simply  to  raise  its
effective temperature in this region. The mean occupation number  $\<n\>$  does
not depend on the dimensionless pumping rate $N$ (for sufficiently large $N$).

The  variance  is

\begin{formula}{}
\sigma_n^2=\<n^2\>-\<n\>^2=\<n\>(1+\<n\>)=
\frac{(1+n_b)(n_b+\theta^2)}{(1-\theta^2)^2}~~,
\end{formula}

\noi and  the   first   non-leading
eigenvector is easily shown to be

\begin{formula}{BoltzEigen}
u_n=\frac{n-\<n\>}{\sigma_n}~~,
\end{formula}

\noi which indeed has the form of  a  univariate  variable.  The  corresponding
eigenvalue is found from Eq.~(\ref{Eigen1})  $\lambda_1=1-\theta^2$, or

\begin{formula}{}
\xi=\inv{1-\theta^2}~~.
\end{formula}

\noi Thus the correlation length diverges at $\theta=1$ (for $N\to\infty$).

\subsection{First critical point: $\theta=1$}

Around the critical point at $\theta=1$ there is competition between
the linear and quadratic terms in the expansion of the potential for small $x$

\begin{formula}{}
V(x)=x\log\frac{n_b+1}{n_b+\theta^2}
+\frac16x^2\frac{\theta^4}{\theta^2+n_b}+\O(x^3)~~.
\end{formula}

\noi Expanding in $\theta^2-1$ we get

\begin{formula}{}
V(x)=\frac{1-\theta^2}{1+n_b}x+\frac1{6(1+n_b)}x^2+\O\(x^3,(\theta^2-1)^2\)~~.
\end{formula}

\noi  Near  the  critical  point,  \ie\  for  $(1-\theta^2)\sqrt{N}\ll1$,   the
quadratic term dominates, so the average value $\<x\>$ as  well  as  the  width
$\sigma_x$ becomes of $\O(1/\sqrt N)$ instead of $\O(1/N)$.

Let us therefore introduce two scaling variables $r$ and $\alpha$ through

\begin{formula}{Scales}
x=r\sqrt{3(1+n_b)\over N}~,~~
\theta^2-1=\alpha\sqrt{1+n_b\over3N}~~,
\end{formula}

\noi so that the probability distribution in terms of these variables
becomes a Gaussian on the half-line, \ie\

\begin{formula}{}
p^0(r)=\frac1{Z_r}e^{-\frac12(r-\alpha)^2}
\end{formula}

\noi with

\begin{formula}{}
Z_r=\int_0^\infty dr\;e^{-\frac12(r-\alpha)^2}
=\sqrt{\frac\pi2}\(1+\erf\(\frac\alpha{\sqrt2}\)\)~~.
\end{formula}

\noi From this we obtain

\begin{formula}{}
\<r\>=\alpha+\frac{d\log Z_r}{d\alpha}~,~~~
\sigma^2_r=\frac{d\<r\>}{d\alpha}~~.
\end{formula}

\noi For $\alpha=0$ we have explicitly

\begin{formula}{}
\<x\>=\sqrt{12(1+n_b)\over \pi N}~,~~
\sigma^2_x=\frac{6(n_b+1)}{N}\(\frac12-\frac1\pi\)~~.
\end{formula}

This leads to the following equation for $u(r)$

\begin{formula}{Critical}
\rho u=r(r-\alpha)\frac{du}{dr}-\frac{d}{dr}\left[r\frac{du}{dr}\right]
{}~~,
\end{formula}

\noi where

\begin{formula}{}
\rho=\lambda\sqrt{\frac{3N}{1+n_b}}=\left\<r\(\frac{du}{dr}\)^2\right\>_0
{}~~.
\end{formula}

\noi This eigenvalue problem has no simple solution.

We know, however, that $u(r)$ must change sign once, say  at  $r=r_0$.  In  the
neighbourhood of the sign change we have $u\simeq r-r_0$ and, inserting  this
into
(\ref{Critical}) we get $r_0=(\alpha+\sqrt{4+\alpha^2})/2$ and
$\rho=\sqrt{4+\alpha^2}$ such that

\begin{formula}{CritXi}
\xi=\sqrt{\frac{3N}{(1+n_b)(4+\alpha^2)}}~~.
\end{formula}

\subsection{Maser phase: $1<\theta<\theta_1\simeq4.603$}

In the region above the transition at $\theta=1$  the  mean  occupation  number
$\<n\>$ grows proportionally with the pumping rate $N$, so in this  region  the
cavity acts as a maser. There is a single minimum of  the  effective  potential
described by the branch $x_0(\theta)$, defined by the  region  $0<\phi<\pi$  in
Eq. (\ref{Extrema}). We find  for  $N\gg1$  to  a  good  approximation  in  the
vicinity of the minimum a Gaussian behaviour

\begin{formula}{}
p^0(x)=\sqrt{\frac{NV''(x_0)}{2\pi}}e^{-\frac N2 V''(x_0)(x-x_0)^2}~~,
\end{formula}

\noi where

\begin{formula}{}
V''(x_0)=\frac{1-q'(x_0)}{x_0(1+n_b)}~~.
\end{formula}

\noi Hence for $(\theta^2-1)\sqrt N\gg1$ we have a
mean value  $\<x\>_0=x_0$  and
variance $\sigma^2_x=1/NV''(x_0)$. To find the  next-to-leading  eigenvalue  in
this case we introduce the scaling variable  $r=\sqrt{NV''(x_0)}(x-x_0)$, which
has zero mean and unit variance for large $N$. Then  Eq.~(\ref{uEigen1})  takes
the form (in the continuum limit $N\to\infty$)

\begin{formula}{}
\lambda u=(1-q'(x_0))\(r\frac{du}{dr}-\frac{d^2u}{dr^2}\)~~.
\end{formula}

\noi This is the differential equation for Hermite polynomials. The eigenvalues
are  $\lambda_n=n(1-q'(x_0))$,  $n=0,1,\ldots$, and grow linearly with $n$.
This may be observed in Fig. \ref{FigFit}. The  correlation
length becomes

\begin{formula}{}
\xi=\inv{1-q'(x_0)}=\inv{1-\phi\cot\phi}\quad\mbox{for $0<\phi<\pi$}~~.
\end{formula}

\noi As in the thermal phase, the correlation length is independent of $N$
(for
large $N$).


\subsection{Mean field calculation}

We shall now use a mean field method to get an expression for  the  correlation
length in both the thermal and maser phases and in the critical region. We find
from the time-dependent probability  distribution  (\ref{Maser})  the
following
{\it exact} equation for the average photon occupation number:

\begin{formula}{}
\frac1\gamma\frac{d\<n\>}{dt}=N\<q_{n+1}\>+n_b-\<n\>
{}~~,
\end{formula}

\noi or with $\Delta x=1/N$

\begin{formula}{}
\frac1\gamma\frac{d\<x\>}{dt}=\<q(x+\Delta x)\>+n_b\Delta x-\<x\>
{}~~.
\end{formula}

\noi We shall ignore the fluctuations of $x$ around its mean value and
simply replace this by

\begin{formula}{}
\frac1\gamma\frac{d\<x\>}{dt}=q(\<x\>+\Delta x)+n_b\Delta x-\<x\>
{}~~.
\end{formula}

\noi This is certainly a good approximation in the limit  of  $N\to\infty$  for
the maser phase because the relative fluctuation $\sigma_x/\<x\>$  vanishes  as
$\O(1/\sqrt N)$ here, but it is of dubious validity in the thermal phase, where
the relative  fluctuations  are  independent  of  $N$.  Nevertheless,  we  find
numerically that the mean field description is  rather  precise  in  the  whole
interval $0<\theta<\theta_1$.

The fixed point $x_0$ of the above equation satisfies the mean field equation

\begin{formula}{}
x_0=q(x_0+\Delta x)+n_b\Delta x
{}~~,\end{formula}

\noi which may be solved in parametric form as

\begin{formulas}{}
x_0&=&\sin^2\phi+n_b\Delta x~~,\\
\theta&=&\displaystyle\frac\phi{\sqrt{\sin^2\phi+(1+n_b)\Delta x}}~~.
\end{formulas}

\noi We notice here that there is a maximum region of existence for any
branch of the solution. The maximum is roughly given by $\theta_k^{\max}
=(k+1)\pi\sqrt{N/(1+n_b)}$.

For small perturbations $\<x\>=x_0+\epsilon$ we find the equation of motion

\begin{formula}{}
\frac1\gamma\frac{d\epsilon}{dt}=-(1-q'(x_0+\Delta x))\epsilon
{}~~,\end{formula}

\noi from which we estimate the leading eigenvalue

\begin{formula}{}
\lambda=1-q'(x_0+\Delta x)
=1-\frac{\phi\sin\phi\cos\phi}{\sin^2\phi+(1+n_b)\Delta x}
{}~~.\end{formula}

\noi
\bort{
In Fig. \ref{FigMeanLambda} this solution is  plotted  as  a  function  of
$\theta$.
}
Notice that $\lambda$ takes negative  values  in  the  unstable  regions  of
$\phi$. This eigenvalue does not vanish at the critical point $\theta=1$  which
corresponds to

\begin{formula}{}
\phi\simeq\phi_0=\(\frac{3(1+n_b)}N\)^{\frac14}
{}~~,
\end{formula}

\noi but only reaches a small value

\begin{formula}{}
\lambda\simeq2\sqrt{\frac{1+n_b}{3N}}~~,
\end{formula}

\noi which agrees exactly with the previously obtained result (\ref{CritXi}).
Introducing the scaling variable  $\alpha$  from  (\ref{Scales})  and  defining
$\psi=(\phi/\phi_0)^2$    we    easily  get

\begin{formulas}{}
\alpha=(\psi^2-1)/\psi~~,\\
r=\psi~~,\\
\rho=(\psi^2+1)/\psi~~,\\
\end{formulas}

\noi and after eliminating $\psi$

\begin{formulas}{}
r=\frac12(\alpha+\sqrt{\alpha^2+4})~~,\\
\rho=\sqrt{4+\alpha^2}~~,\\
\end{formulas}

\noi which agrees with the previously obtained results.

\subsection{The first critical phase:
$4.603\simeq\theta_1<\theta<\theta_2\simeq7.790$}\label{FirstCritical}

We now turn to the first phase in which the effective potential has two  minima
($x_0, x_2$) and a maximum ($x_1$) in between (see Fig. \ref{FigEffPot} in
Section~\ref{s:Extrema}).
In this case there is competition between the two  minima
separated by the barrier and for $N\to\infty$ this barrier makes the relaxation
time to equilibrium  exponentially long.
Hence  we  expect  $\lambda_1$  to  be
exponentially small for large $N$
(see \fig{FigCritic})

\begin{formula}{Asymp}
\lambda_1=Ce^{-\eta N}~~,
\end{formula}

\noi where $C$ and $\eta$ are independent of $N$. It is the  extreme  smallness
of the subleading  eigenvalue  that  allows  us  to  calculate  it  with  high
precision.

For large $N$ the probability  distribution  consists  of  two  well-separated
narrow maxima, each of  which  is  approximately  a  Gaussian.  We  define  the
\apriori\ probabilities for each of the peaks

\begin{formula}{}
P_0=\sum_{0\le x<x_1}\Delta x\; p^0(x)=\frac{Z_0}{Z}~~,
\end{formula}

\noi and

\begin{formula}{}
P_2=\sum_{x_1\le x<\infty}\Delta x\; p^0(x)=\frac{Z_2}{Z}~~.
\end{formula}

\noi The $Z$-factors are

\begin{formula}{}
Z_0=\sum_{x=0}^{x_1}\Delta x\;e^{-NV(x)}
\simeq e^{-NV_0}\sqrt{\frac{2\pi}{NV''_0}}~~,
\end{formula}

\noi and

\begin{formula}{}
Z_2=\sum_{x=x_1}^\infty\Delta x\;e^{-NV(x)}
\simeq e^{-NV_2}\sqrt{\frac{2\pi}{NV''_2}}~~,
\end{formula}

\noi with $Z=Z_0+Z_2$. The probabilities satisfy of course $P_0+P_2=1$ and
we have

\begin{formula}{}
p^0(x)=P_0p^0_0(x)+P_2p^0_2(x)~~,
\end{formula}

\noi where $p^0_{0,2}$ are individual probability distributions with maximum at
$x_{0,2}$.  The  overlap  error  in  these  expressions  vanishes  rapidly  for
$N\to\infty$, because  the  ratio  $P_0/P_2$  either  converges  towards  0  or
$\infty$ for $V_0\ne V_2$. The transition from one peak being  the  highest  to
the other peak being the highest occurs when the two maxima coincide, \ie\  at
$\theta\simeq7.22$  at  $N=10$,  whereas  for  $N=\infty$   it   happens  at
$\theta\simeq6.66$. At this point the correlation length is also maximal.

Using this formalism, many quantities may be evaluated in the limit
of large $N$. Thus for example

\begin{formula}{}
\<x\>_0=P_0x_0+P_2x_2~~,
\end{formula}

\noi and

\begin{formula}{}
\sigma_x^2=\<(x-\<x\>_0)^2\>_0=\sigma_0^2P_0+\sigma_2^2P_2
+(x_0-x_2)^2P_0P_2 ~~.
\end{formula}

\noi Now there is no direct relation between the variance and  the  correlation
length.

Consider now the expression (\ref{pEigen}),
which shows that since $\lambda_1$
is exponentially small we have an essentially constant $J_n$, except  near  the
maxima of the probability distribution, \ie\ near the minima of the  potential.
Furthermore since  the  right  eigenvector  of  $\lambda_1$  satisfies  $\sum_x
p(x)=0$, we have $0=J(0)=J(\infty)$ so that

\begin{formula}{}
J(x)\simeq
\left\{\begin{array}{ll}
0&0<x<x_0~~,\\
J_1&x_0<x<x_2~~,\\
0&x_2<x<\infty~~.\\
\end{array}\right.
\end{formula}

\noi This expression is more accurate away from the minima of the potential,
$x_0$ and $x_2$.

Now it follows  from  Eq.~(\ref{uEq})  that  the  left  eigenvector  $u(x)$  of
$\lambda_1$ must be constant, except near the minimum $x_1$ of the  probability
distribution, where the derivative could be sizeable.
So we conclude that $u(x)$
is constant away from the maximum of the potential. Hence we must approximately
have

\begin{formula}{}
u(x)\simeq
\left\{\begin{array}{ll}
u_0&0<x<x_1~~,\\
u_2&x_1<x<\infty~~.\\
\end{array}\right.
\end{formula}

\noi This expression is more accurate away from the maximum of the potential.

We may now relate the values of $J$ and $u$  by summing Eq.~(\ref{pEigen})
from $x_1$ to infinity

\begin{formula}{}
J_1=J(x_1)=\lambda_1\sum_{x=x_1}^\infty \Delta x\;p^0(x)u(x)
\simeq \lambda_1 P_2u_2~~.
\end{formula}

\noi From Eq.~(\ref{uEq}) we get by summing over the interval between the
minima

\begin{formula}{}
u_2-u_0=\frac{NJ_1}{1+n_b}\sum_{x=x_0}^{x_2}\Delta x  \frac1{xp^0(x)}~~.
\end{formula}

\noi The inverse probability distribution has for $N\to\infty$ a sharp
maximum at the maximum of the potential. Let us define

\begin{formula}{}
Z_1=\sum_{x=x_0}^{x_2} \Delta x\;\frac1xe^{NV(x)}
\simeq \frac1{x_1}e^{NV_1}\sqrt{\frac{2\pi}{N(-V''_1)}}~~.
\end{formula}

\noi Then we find

\begin{formula}{}
u_2-u_0=\frac{NZZ_1J_1}{1+n_b}=\frac{N\lambda_1 Z_1Z_2u_2}{1+n_b}~~.
\end{formula}

\noi But $u(x)$ must be univariate, \ie\

\begin{formulas}{}
u_0P_0+u_2P_2=0~~,\\
u_0^2P_0+u_2^2P_2=1~~,\\
\end{formulas}

\noi from which we get

\begin{formulas}{}
u_0=-\sqrt{\displaystyle{\frac{P_2}{P_0}}}
=-\sqrt{\displaystyle{\frac{Z_2}{Z_0}}}~~,\\\\
u_2=\sqrt{\displaystyle{\frac{P_0}{P_2}}}
=\sqrt{\displaystyle{\frac{Z_0}{Z_2}}}~~.\\
\end{formulas}

\noi Inserting the above solution we may solve for $\lambda_1$

\begin{formula}{LambdaCrit}
\lambda_1=\frac{1+n_b}{N}\frac{Z_0+Z_2}{Z_0Z_1Z_2}~~,
\end{formula}

\noi or more explicitly

\begin{formula}{glorylam}
\lambda_1=\frac{x_1(1+n_b)}{2\pi}\sqrt{-V_1''}
\(\sqrt{V_0''}e^{-N(V_1-V_0)}+\sqrt{V_2''}e^{-N(V_1-V_2)}\)~~.
\end{formula}

\noi
Finally we may read off the coefficients $\eta$ and $C$ from Eq.~(\ref{Asymp}).
We get

\begin{formula}{Eta}
\eta=\left\{\begin{array}{ll}
V_1-V_0&{\rm for}~~ V_0>V_2~~,\\
V_1-V_2&{\rm for}~~ V_2>V_0~~,\\
\end{array}\right.
\end{formula}

\noi and

\begin{formula}{}
C=\frac{x_1(1+n_b)}{2\pi}\sqrt{-V_1''}
\left\{\begin{array}{ll}
\sqrt{V_0''}&{\rm for}~~ V_0>V_2~~,\\
\sqrt{V_2''}&{\rm for}~~ V_2>V_0~~.\\
\end{array}\right.
\end{formula}

This expression is nothing  but the result of a barrier  penetration  of  a
classical statistical process \cite{Schuss80}. We have derived it in
detail in order to get all the coefficients right.

\begin{figure}[htb]
\unitlength=1mm
\begin{picture}(140,90)(0,0)
\includegraphics{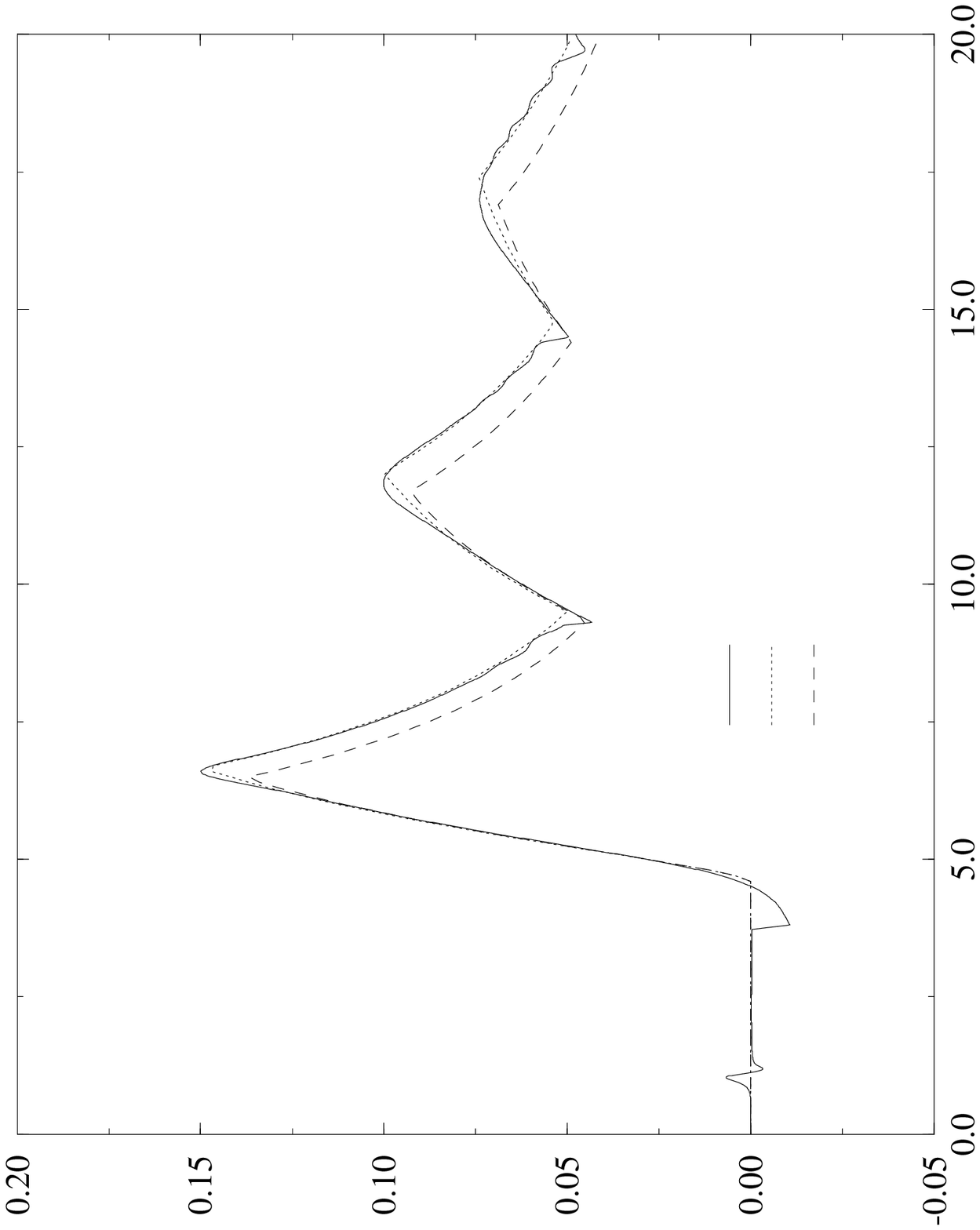}
\put(75,0){\small$\theta$}
\put(10,50){\small$\eta$}
\put(78,25){\small$\log(\xi_{90}/\xi_{70})/20$}
\put(78,20.5){\small Barrier from $V(x)$}
\put(78,16){\small Barrier from Fokker--Planck}
\end{picture}
\figcap{Comparing the barrier height from the potential $V(x)$ with  the  exact
correlation length and the barrier from an approximate Fokker--Planck formula.}
\label{f-barr}
\end{figure}

It  is  interesting  to  check  numerically  how  well  \eq{glorylam}  actually
describes the correlation length.  The  coefficient  $\eta$  is  given
by  Eq.~(\ref{Eta}), and we have numerically computed
the  highest  barrier  from  the
potential $V(x)$ and compared it with an exact calculation in \fig{f-barr}. The
exponent $\eta$ is extracted by comparing two values of the correlation length,
$\xi_{70}$ and $\xi_{90}$, for large values  of  $N$  (70  and  90),
  where  the
difference in the prefactor $C$ should be unimportant.  The  agreement  between
the two calculations is excellent  when  we  use  the  exact  potential.  As  a
comparison  we  also  calculate  the  barrier  height  from  the  approximative
potential     in     the      Fokker--Planck      equation      derived      in
\cite{Filipowicz86,Filipowicz86a}. We find a  substantial  deviation  from  the
exact   value   in    that    case.    It    is    carefully    explained    in
\cite{Filipowicz86,Filipowicz86a} why the Fokker--Planck  potential  cannot  be
expected to give a quantitatively correct  result  for  small  $n_b$.  The
exact
result (solid line) has some  extra  features  at  $\theta=1$  and  just  below
$\theta=\theta_1\simeq 4.603$, due to finite-size effects.

When  the  first  subleading  eigenvalue  goes  exponentially   to   zero,   or
equivalently the correlation length grows exponentially, it  becomes  important
to know the density of eigenvalues. If there is an accumulation of  eigenvalues
around 0, the long-time correlation cannot be  determined  by  only  the  first
subleading  eigenvalue.  It  is  quite  easy  to  determine  the  density   of
eigenvalues simply by computing them numerically.

\begin{figure}[htb]
\unitlength=1mm
\begin{picture}(140,100)(0,0)
\includegraphics{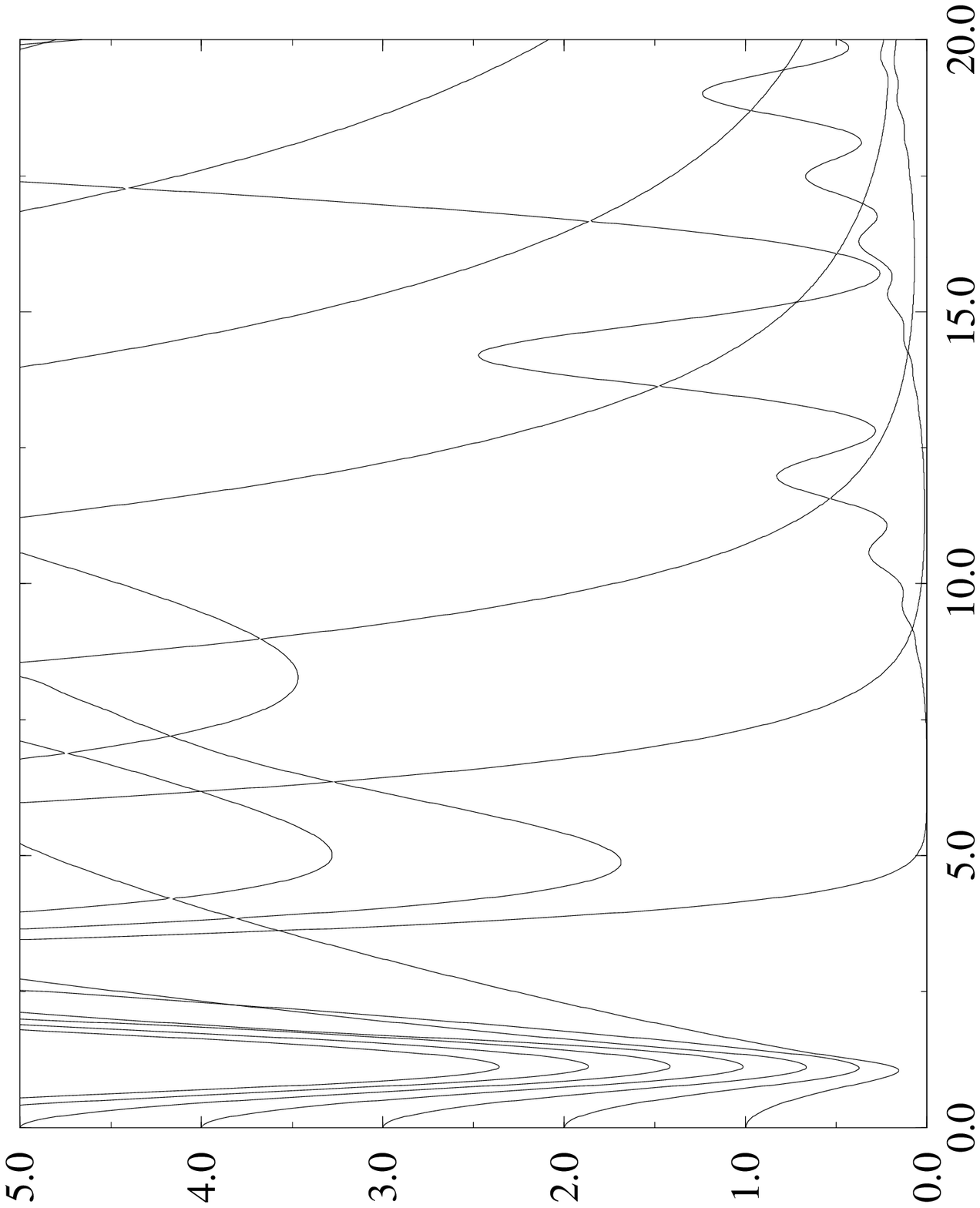}
\put(85,3){\small$\theta$}
\put(10,53){\small$\lambda_{1-7}$}
\end{picture}
\figcap{The first seven subleading eigenvalues for $N=50$ and $n_b=0.15$.}
\label{f-lam0-7}
\end{figure}

In \fig{f-lam0-7} we show the first seven subleading eigenvalues for $N=50$ and
$n_b=0.15$. It is clear that at the first critical point after the maser  phase
($\theta=\theta_1$) there is only one eigenvalue going to zero.  At  the  next
critical phase $(\theta=\theta_2$) there is one more  eigenvalue  coming  down,
and so on. We find that there is only one exponentially  small  eigenvalue  for
each new minimum in the  potential,  and  thus  there  is  no  accumulation  of
eigenvalues around 0.


\def\st{\sigma_\theta}

\section{Effects of velocity fluctuations}
\seqnoll
\label{Spread}

The time it takes an atom to  pass  through  the  cavity  is  determined  by  a
velocity filter in front of the cavity. This filter is not perfect  and  it  is
relevant to investigate what a spread  in  flight  time  implies  for  the
statistics of the interaction between cavity  and  beam.  To  be  specific,  we
consider the flight time as an independent stochastic variable.  Again,  it  is
more convenient to work with the rescaled variable $\theta$, and we denote  the
corresponding stochastic variable by $\vartheta$.  In  order  to  get  explicit
analytic results we choose the following probability distribution for  positive
$\vartheta$

\begin{formula}{distr}
    f(\vartheta,\alpha,\beta)=\frac{\beta^{\alpha+1}}{\Gamma(\alpha+1)}
    \vartheta^\alpha
    e^{-\beta\vartheta}~~,
\end{formula}

\noi  with  $\beta=\theta/\st^2$   and   $\alpha=\theta^2/\st^2-1$,   so   that
$\<\vartheta\>=\theta$ and $\<(\vartheta-\theta)^2\>=\st^2$. Other choices  are
possible, but are not expected to change the overall qualitative  picture.  The
discrete master equation (\ref{stat1}) for the equilibrium distribution can  be
averaged to yield

\begin{formula}{avemast}
    \< p(t+T)\>=e^{-\gamma L_C T}\< M(\vartheta)\> \< p(t)\>~~,
\end{formula}

\noi The factorization  is  due  to  the  fact  that  $p(t)$  only  depends  on
$\vartheta$ for the preceding atoms,  and  that  all  atoms  are  statistically
independent.      The       effect       is       simply       to       average
$q(\vartheta)=\sin^2(\vartheta\sqrt{x})$ in $M(\vartheta)$, and we get

\begin{formula}{aveq}
    \< q\>=\inv2 \left[1-
      \left(1+\frac{4x\st^4}{\theta^2}\right)^
      {-\frac{\theta^2}{2\st^2}}
      \cos\left(\frac{\theta^2}{\st^2}
        \arctan\left(\frac{2\sqrt{x}\st^2}{\theta}\right)\right)\right]~~.
\end{formula}

\noi This averaged form of $q(\theta)$, which depends on  the  two  independent
variables $\theta/\st$ and $\theta \sqrt{x}$, enters in  the  analysis  of  the
phases in exactly the same way as before. In the limit $\st\goto 0$  we  regain
the original $q(\theta)$, as we should. For  very  large  $\st$  and  fixed
$\theta$, and $\<q\>$ approaches zero.

\subsection{Revivals and prerevivals}
\label{ss:prerev}

The phenomenon of quantum revival is an essential  feature  of  the  microlaser
system (see \cite{Meystre74}--\cite{Filipowicz86b}, and
\cite{Averbukh89a}--\cite{Fleischhauer93}).
The  revivals  are  characterized  by the
reappearance  of  strongly   oscillating
structures in the excitation probability of an outgoing atom which is given  by
Eq.~(\ref{SingleP}):

\begin{formula}{cp}
\P(+)={u^0}^TM(+)p^0=\sum_n (1-q_{n+1}(\theta)) p^0_n~~,
\end{formula}

\noi where $p^0_n$ is the photon distribution (\ref{Equilibrium}) in the cavity
before the atom enters. Revivals occur when there is a resonance  between  the
period in $q_n$ and the  discreteness  in  $n$  \cite{Fleischhauer93}.  If  the
photon distribution in the cavity has a sharp peak at $n=n_0$ with  a  position
that does not change appreciably when $\theta$ changes, as for  example  for  a
fixed Poisson distribution, then it is easy  to  see  that  the  first  revival
becomes pronounced in the region of $\theta_{\rm  rev}\simeq  2\pi\sqrt{n_0N}$.
For the equilibrium distribution without any spread in the velocities we do not
expect any dramatic signature of revival, the reason being that  the  peaks  in
the equilibrium distribution $p^0_n(\theta)$ move  rapidly  with  $\theta$.  In
this context it is also natural to study the short-time correlation between two
consecutive atoms, or the probability of finding two consecutive atoms  in  the
excited level \cite{PaulRichter91}. This quantity is given by

\begin{formulas}{}
\P_0(+,+)&=&{u^0}^TM(+)(1+L_C/N)^{-1}M(+)p^0\\
&=&\displaystyle{\sum_{n,m}}(1-q_{n+1}(\theta))(1+L_C/N)^{-1}_{nm}
(1-q_{m+1}(\theta)) p^0_m~~,
\end{formulas}

\begin{figure}[p]
\unitlength=1mm
\begin{picture}(140,90)(0,0)
\includegraphics{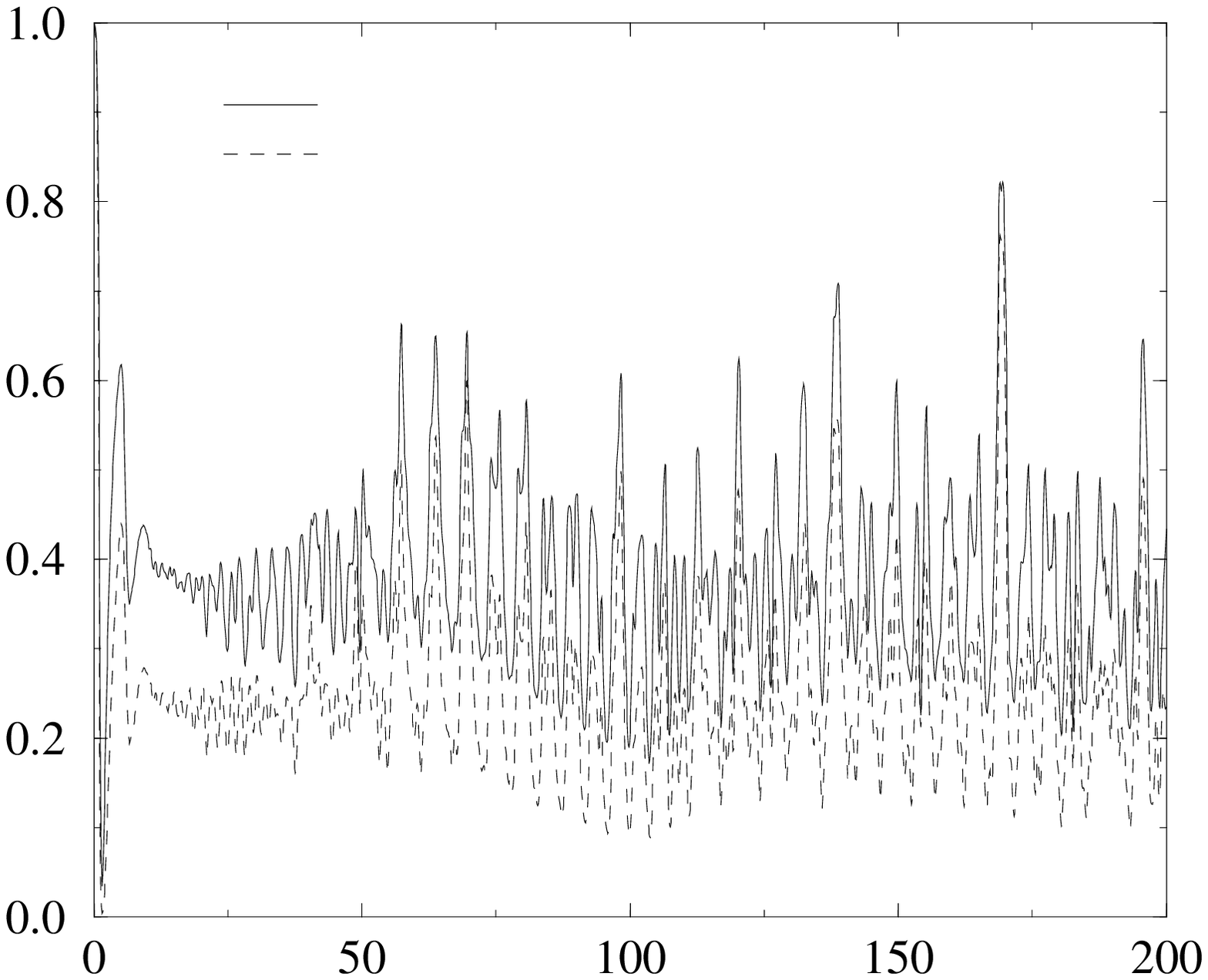}
\put(50,86){\small$\P(+)$}
\put(50,81){\small$\P_0(+,+)$}
\end{picture}

\begin{picture}(140,90)(0,0)
\includegraphics{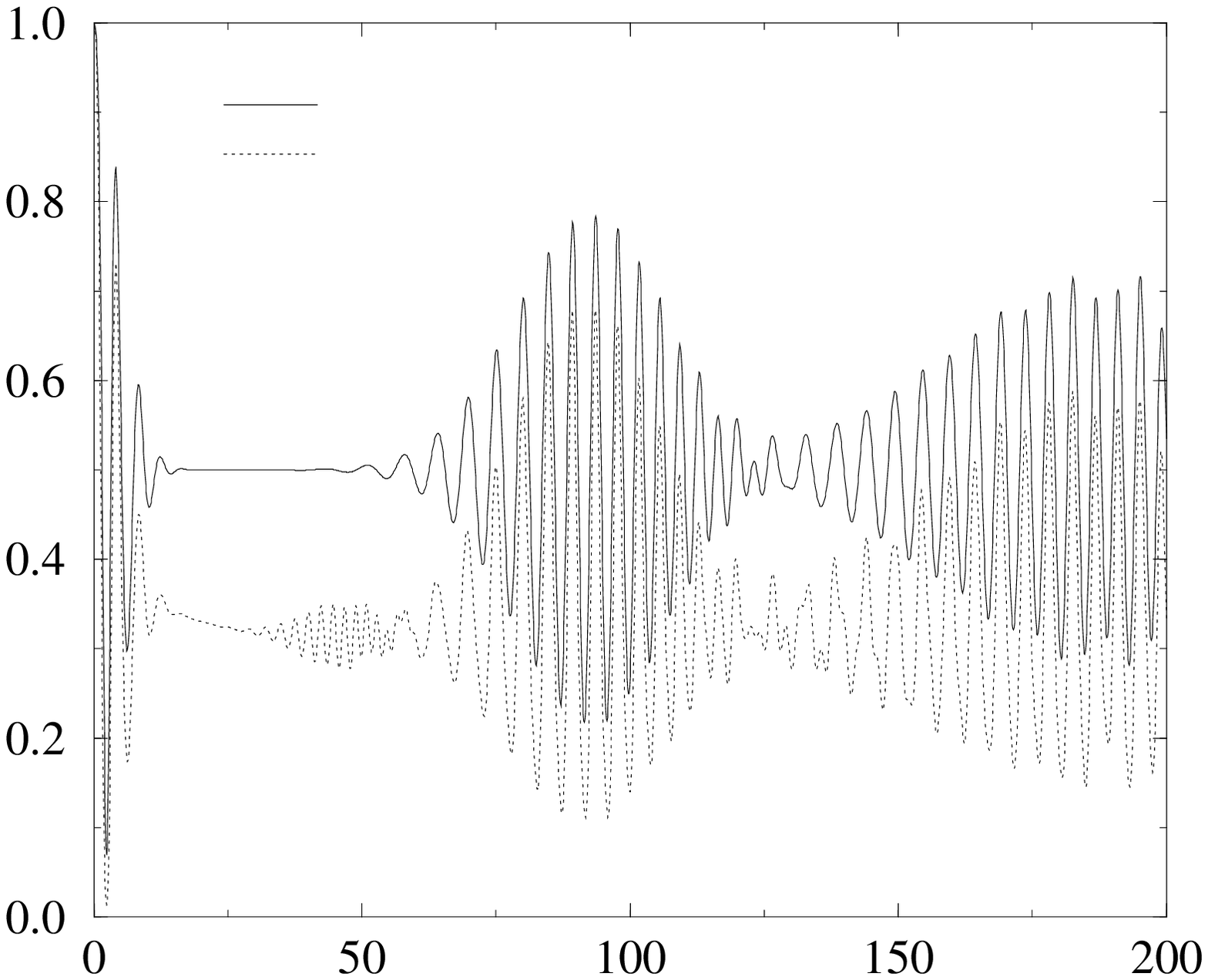}
\put(75,2){\small$\theta$}
\put(50,86){\small$\P(+)$}
\put(50,81){\small$\P_0(+,+)$}
\end{picture}

\figcap{Upper graph: Probabilities of finding one atom,
or two consecutive ones,  in
the excited state. The flux is given  by  $N=20$  and  the  thermal  occupation
number is $n_b=0.15$. The curves show no evidence for the resonant behaviour of
revivals. Lower graph: Presence of  revival  resonances  in  equilibrium  after
averaging the photon distribution over $\theta$.  The  same  parameters  as  in
\fig{f:rev0} are used but the variance in $\theta$ is now given by $\st^2=10$.}
\label{f:rev0}
\end{figure}

\noi defined in Eq.~(\ref{kProb}). In Appendix \ref{dampingmatrix} we  give  an
analytic  expression  for  the  matrix   elements   of   $(1+L_C/N)^{-1}$.   In
\fig{f:rev0} we present $\P(+)$ and $\P_0(+,+)$ for typical values of  $N$  and
$n_b$.

If we on the other hand smear out the equilibrium distribution sufficiently  as
a function of $\theta$, revivals will again appear. The experimental  situation
we envisage is that the  atoms  are  produced  with a
certain  spread  in  their
velocities. The statistically averaged stationary photon  distribution  depends
on the spread. After the  passage  through  the  cavity  we  measure  both  the
excitation level and the speed of the atom. There is thus no averaging  in  the
calculation of $\P(+)$ and $\P_0(+,+)$, but these quantities now also depend on
the actual value $\vartheta$ for each atom. For  definiteness  we  select  only
those atoms that fall in a narrow range around the average value  $\theta$,  in
effect putting in a sharp velocity filter after the interaction. The result for
an averaged photon distribution is presented  in  \fig{f:rev0}  (lower  graph),
where clear signs of revival are found. We also  observe  that  in  $\P_0(+,+)$
there are {\em prerevivals}, occurring
for a value of $\theta$ half as large  as
for the usual revivals. Its origin is obvious since in  $\P_0(+,+)$  there  are
terms containing $q_n^2$ that vary with the double of the frequency of $q_n$.

\subsection{Phase diagram}
\label{ss:avephd}

The different phases discussed in Section \ref{Phasestructure} depend  strongly
on the structure of the effective potential. Averaging over $\theta$ can easily
change this structure and the phases. For instance, averaging with large  $\st$
would typically wash out some of the minima and lead to  a  different  critical
behaviour. We shall determine a two-dimensional phase diagram in the parameters
$\theta$ and $\st$ by finding the lines where new minima occur  and  disappear.
They are determined by the equations

\begin{formulas}{phdeq}
    \< q\>=x~~,\\
    {\displaystyle\frac{d\< q\>}{dx}}=1~~.
\end{formulas}

\begin{figure}[htb]
\unitlength=1mm
\begin{picture}(140,110)(0,0)
\includegraphics{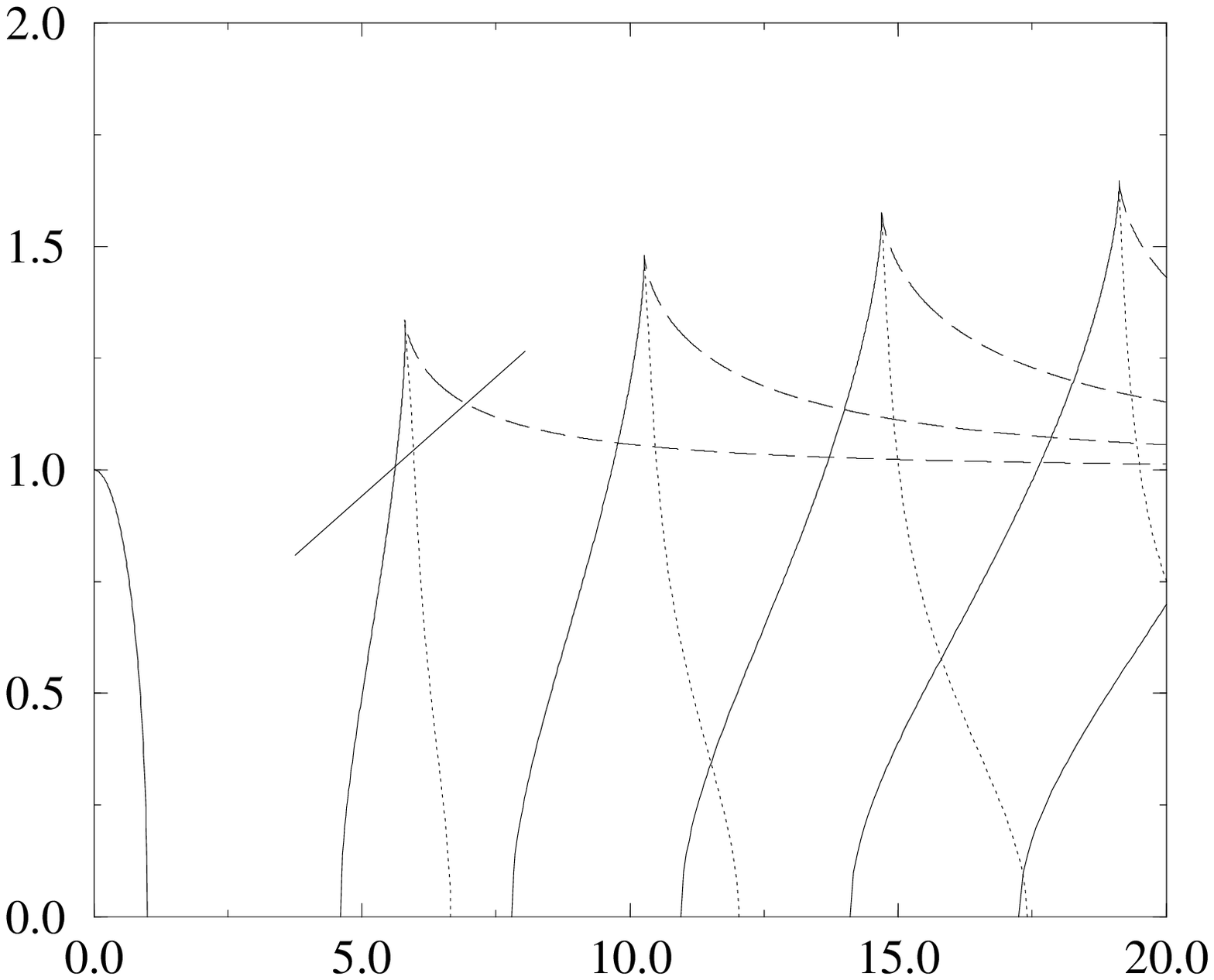}
\end{picture}
\put(-60,5){$\theta$}
\put(-130,70){$\st$}
\put(-85,80){\small $a$}
\put(-100,50){\small $b$}
\put(-71,76){\small $c$}
\figcap{Phase diagram in the $\theta$--$\st$ plane. The  solid  lines  indicate
where new minima in the effective potential emerge. In the lower  left  corner
there is only one minimum at $n=0$, this is the  thermal  phase.  Outside  that
region there is always a minimum for non-zero $n$ implying that the cavity acts
as a maser. To the right of the solid line starting at  $\theta\simeq4.6$,  and
for not too large $\st$, there are two or more minima and thus the  correlation
length grows exponentially with the flux. For increasing $\st$ minima disappear
across the dashed lines, starting with those at small  $n$.  The  dotted  lines
show where the two lowest minima are equally deep.}
\label{f:avephd}
\end{figure}

The phase boundary between the thermal and the maser phase
 is determined by the effective potential  for
small  $x$.  The   condition   $\theta^2=1$   is   now   simply   replaced   by
$\<\vartheta^2\>=\theta^2+\st^2=1$, which also follows from the  explicit  form
of $\<q\>$ in \eq{aveq}. The transitions from the maser phase to  the
critical phases are determined numerically and presented in \fig{f:avephd}. The
first line starting from $\theta\simeq4.6$ shows where the  second  minimum  is
about to form, but exactly on this line it is only an inflection point. At  the
point $a$ about $\st\simeq1.3$ it disappears,  which  occurs  when  the  second
minimum fuses with the first minimum. From the cusp at point $a$ there is a new
line (dashed) showing where the first  minimum  becomes  an  inflection  point.
Above the cusp at point $a$ there is only one minimum.  Going  along  the  line
from point $b$ to $c$ we thus first have one minimum,  then  a  second  minimum
emerges, and finally the first minimum disappears before we  reach  point  $c$.
Similar things happens at the other cusps,
 which represent the fusing points for
other minima. Thus, solid lines show where a new minimum emerges for large  $n$
($\sim N$) as $\theta$ increases, while  dashed  lines  show  where  a  minimum
disappears for small $n$ as $\st$ increases. We have also indicated (by  dotted
lines) the first-order maser transitions where  the  two  dominant  minima  are
equally deep. These are the lines where $\xi$ and  $Q_f$  have  peaks  and
$\<n\>$ makes a discontinuous jump.

\section{Finite-flux effects}
\label{finite}
\seqnoll

So far, we have mainly discussed characteristics of the large flux limit. These
are  the  defining   properties   for   the   different   phases   in   Section
\ref{Phasestructure}. The parameter that controls finite flux  effects  is  the
ratio between the period of oscillations in the potential and the size  of  the
discrete steps in $x$. If $q=\sin^2(\theta\sqrt{x})$ varies slowly over $\Delta
x=1/N$, the continuum limit is usually a good approximation, while  it  can  be
very poor in the opposite case. In the discrete case there exist,  for  certain
values of $\theta$, states that cannot be pumped above  a  certain  occupation
number since $q_n=0$ for that level. This effect is not seen in  the  continuum
approximation.   These   states    are    called    {\em    trapping    states}
\cite{Filipowicz86c} and  we  discuss  them  and  their  consequences  in  this
section.

The continuum approximation starts breaking down for small photon numbers  when
$\theta\simgeq  2\pi\sqrt{N}$,  and  is  completely  inappropriate   when   the
discreteness is manifest for  all  photon  numbers lower  than  $N$,  \ie\  for
$\theta\simgeq    2\pi    N$.    In    that    case     our     analysis     in
Section~\ref{Phasestructure} breaks down and the system  may  occasionally,
  for
certain values of $\theta$, return to a non-critical phase.


\subsection{Trapping states}\label{Trapping}

The  equilibrium  distribution  in  \eq{Equilibrium}  has  peculiar  properties
whenever $q_m=0$ for some value of $m$, in particular when
$n_b$ is  small,  and
dramatically    so    when    $n_b=0$.    This    phenomenon    occurs     when
$\theta=k\pi\sqrt{N/m}$ and is called a trapping state.  When  it  happens,  we
have $p_n=0$ for all $n\geq m$ (for $n_b=0$). The physics behind  this  can  be
found in \eq{Pump},
 where $M(-)$ determines the pumping of  the  cavity  by  the
atoms. If $q_m=0$ the cavity cannot be pumped above  $m$  photons  by  emission
from the  passing  atoms.  For  any
non-zero  value  of  $n_b$  there  is  still  a
possibility for thermal fluctuation above $m$ photons and $p_n\neq 0$ even  for
$n\geq m$. The effect of trapping is lost in  the  continuum  limit  where  the
potential  is  approximated   by   Eq.~(\ref{ContPot}).   Some   experimental
consequences of trapping states  were  studied  for  very  low  temperature  in
\cite{Meystre88} and it was stated that  in  the  range  $n_b=0.1$--1.0  no
experimentally measurable effects were present. We, however,  show  below  that
there are clear signals of trapping states in the correlation length  even  for
$n_b=1.0$.


\subsection{Thermal cavity revivals}
\label{cavrev}

Due to the trapping states, the cavity  may  revert  to  a  statistical  state,
resembling the thermal state at $\theta=0$,  even  if  $\theta>0$.  By  thermal
revival we mean that the state of the cavity returns to the $\theta=0$  thermal
state for other values of $\theta$. Even if the equilibrium state for  non-zero
$\theta$ can resemble a thermal state,
 it does not at all mean that the dynamics
at that value of $\theta$ is similar to what it is  at  $\theta=0$,  since  the
deviations  from  equilibrium  can  have  completely  different  properties.  A
straightforward measure of the deviation  from  the  $\theta=0$  state  is  the
distance in the $L^2$ norm

\begin{formula}{dist}
    d_{L^2}(\theta)=\left(\sum_{n=0}^\infty
    [p_n(0)-p_n(\theta)]^2\right)^{1/2}~~.
\end{formula}

\begin{figure}[htb]
\unitlength=1mm
\begin{picture}(140,90)(0,0)
\includegraphics{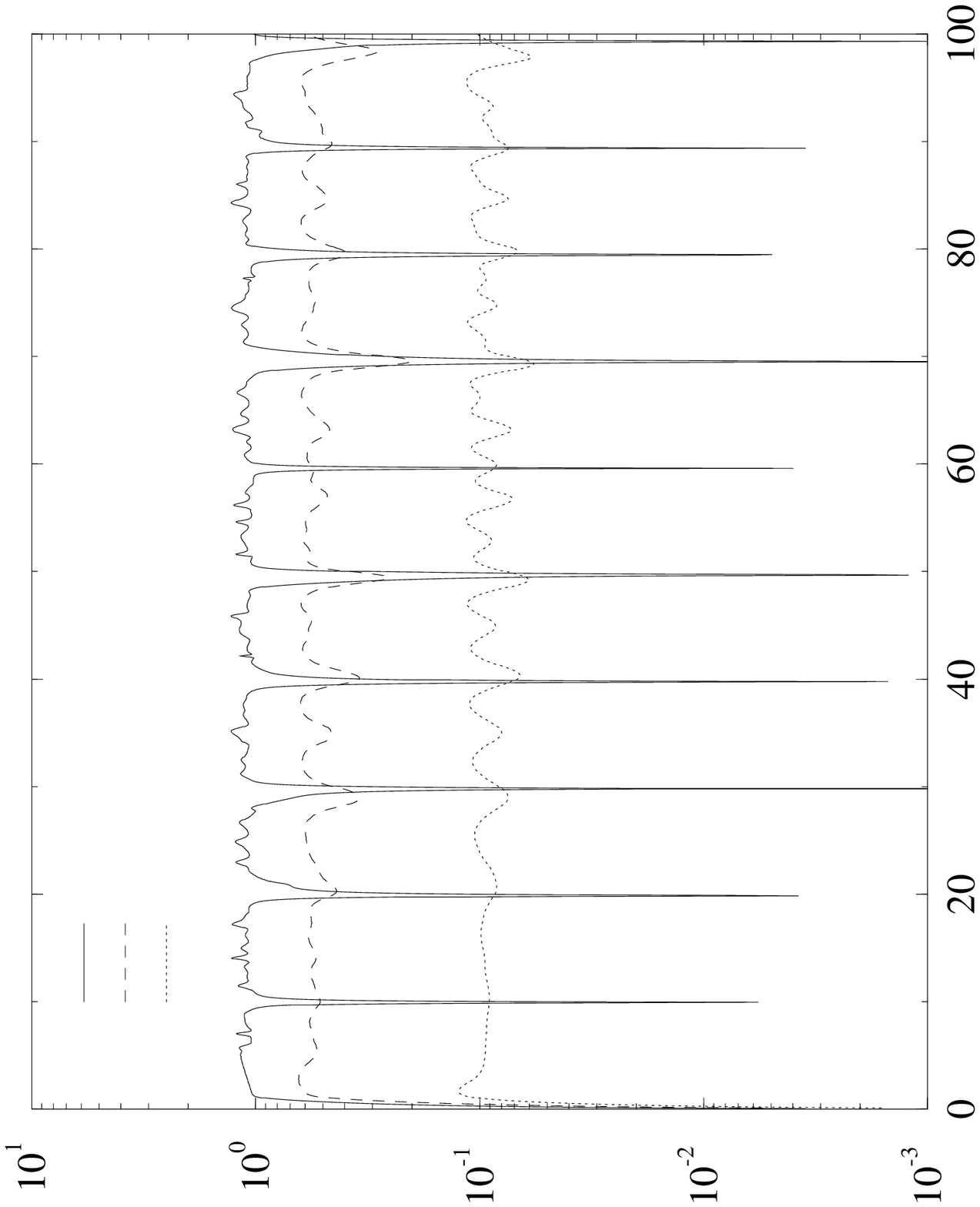}
\put(80,-1){$\theta$}
\put(9,53){$d_{L^2}$}
\put(48,83){\footnotesize$n_b=0.0001$}
\put(48,79){\footnotesize$n_b=1.0$}
\put(48,75){\footnotesize$n_b=10.0$}
\end{picture}
\figcap{Distance between the initial probability
distribution $p_n(0)$ and $p_n(\theta)$ measured by $d_{L^2}(\theta)$
in \eq{dist}.}
\label{f-dist}
\end{figure}

\noi In \fig{f-dist} we exhibit $d_{L^2}(\theta)$ for $N=10$ and several
values of $n_b$.

For small  values  of  $n_b$  we  find  cavity revivals  at  all  multiples  of
$\sqrt{10}\pi$,
  which  can  be  explained  by  the   fact   that   $\sin(\theta
\sqrt{n/N})$ vanishes for $n=1$ and $N=10$ at those points, \ie\ the cavity  is
in a trapping state. That implies  that  $p_n$  vanishes  for  $n\geq  1$  (for
$n_b=0$) and thus there are no photons in the  cavity.  For  larger  values  of
$n_b$ the trapping is less efficient and the  thermal  revivals  go  away.

\begin{figure}[htb]
\unitlength=1mm
\begin{picture}(140,60)(0,0)
\includegraphics{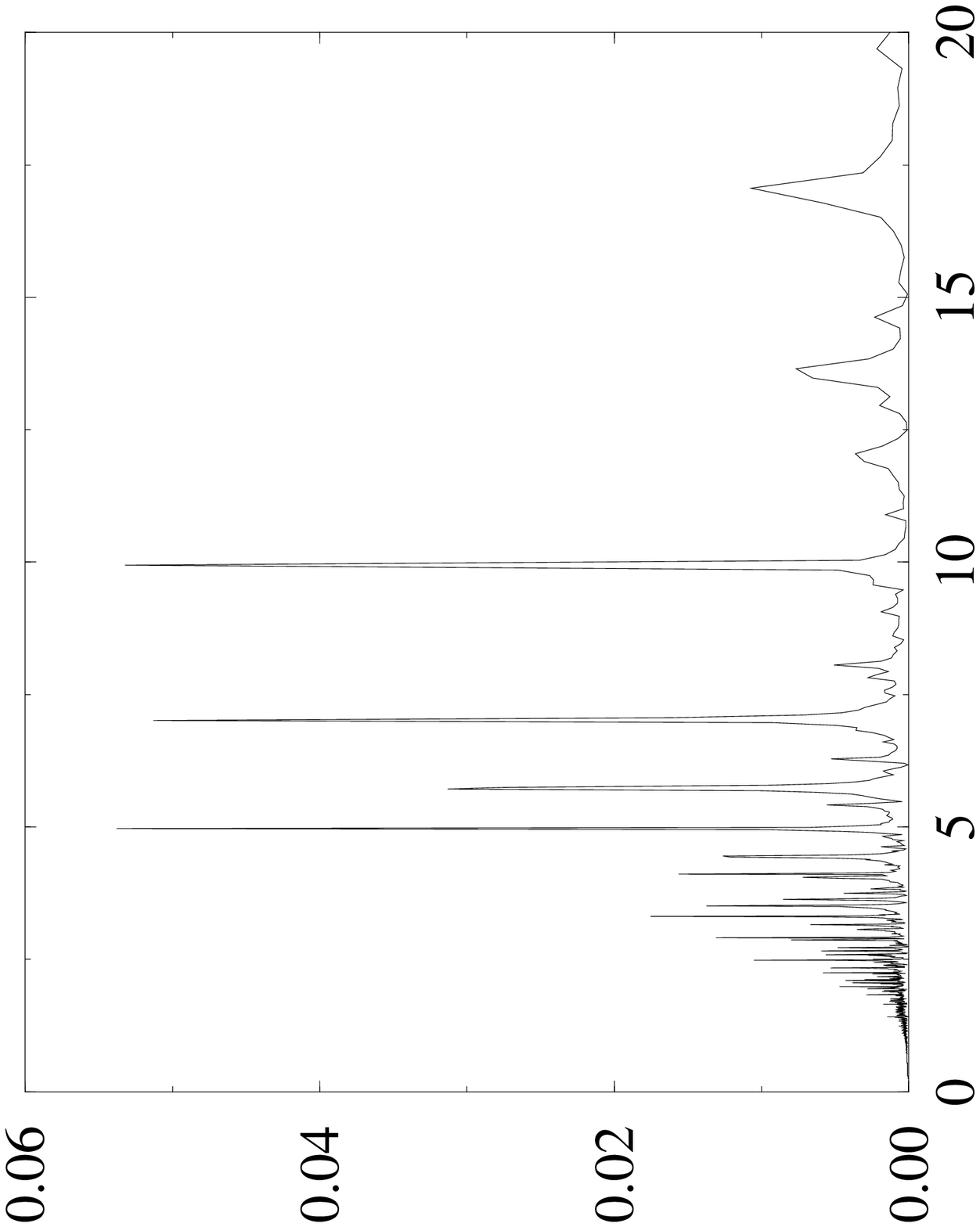}
\put(9,30){\small$\abs{\cF d_{L^2}(\theta)}$}
\end{picture}

\begin{picture}(140,70)(0,0)
\includegraphics{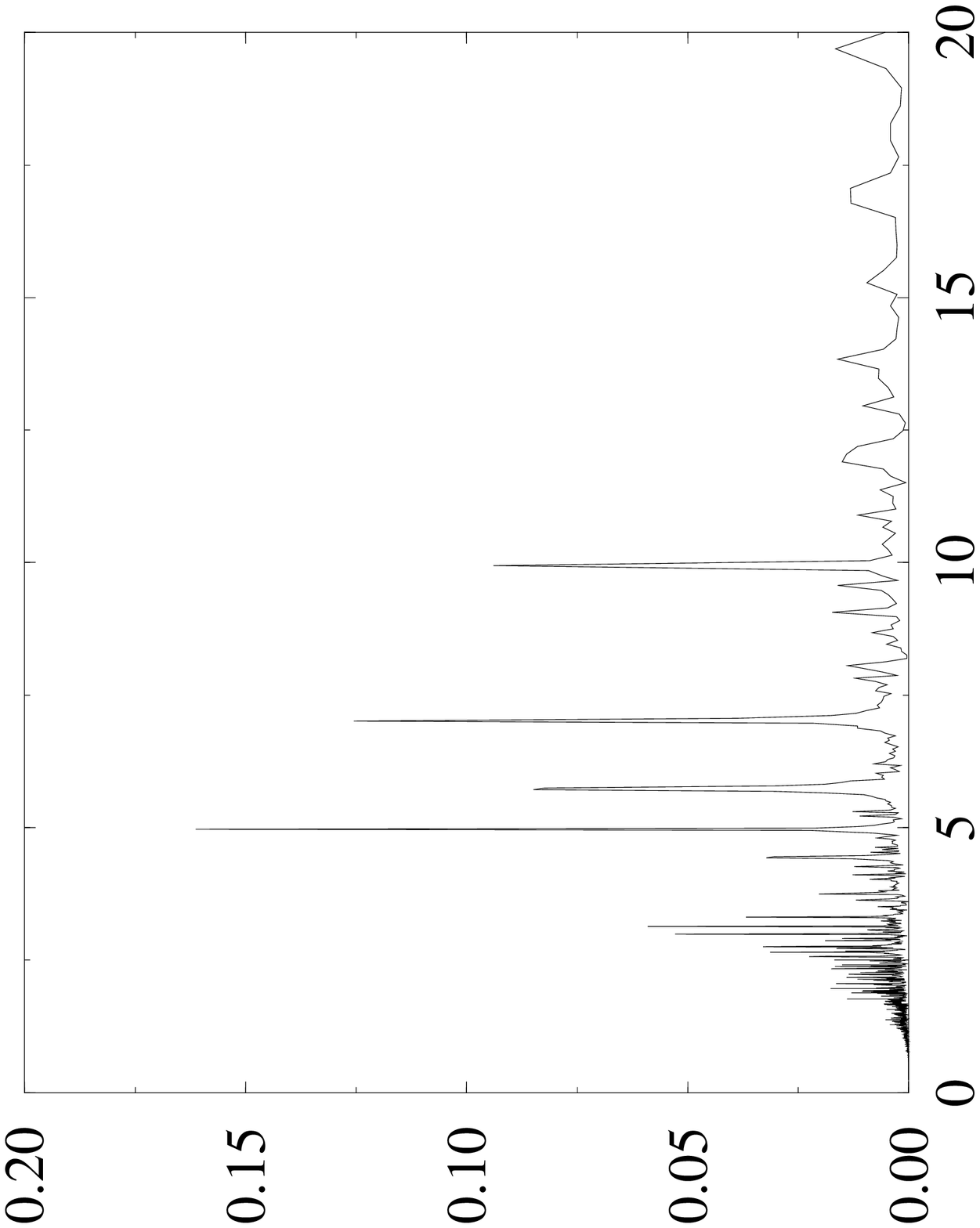}
\put(80,5){\small$\Delta\theta$}
\put(9,48){\small$\abs{\cF \xi(\theta)}$}
\end{picture}
\figcap{Amplitudes of Fourier modes  of  $d_{L^2}(\theta)$  (upper  graph)  and
$\xi(\theta)$ (lower graph) as functions of periods using $N=10$, $n_b=1.0$
and
scanning $0<\theta<1024$. There are pronounced peaks at the values of  trapping
states: $\Delta\theta=\pi \sqrt{N/n}$.}
\label{f-FTFxid}
\end{figure}

Going to much larger values of $\theta$ we can start to look for  periodicities
in the fluctuations in $d_{L^2}(\theta)$. In \fig{f-FTFxid}  (upper  graph)  we
present the spectrum of periods occurring in $d_{L^2}(\theta)$ over  the  range
$0<\theta<1024$.

Standard  revivals  should  occur  with   a   periodicity   of   $\Delta\theta=
2\pi\sqrt{\<n\>}$, which is typically between 15 and 20,
 but there  are  hardly  any  peaks  at
these
values. On the other hand, for periodicities corresponding to trapping  states,
\ie\ $\Delta\theta=\pi\sqrt{10/n}$, there are very  clear  peaks,  even  though
$n_b=1.0$, which is a relatively large value.

\begin{figure}[htb]
\unitlength=1mm
\begin{picture}(140,100)(0,0)
\includegraphics{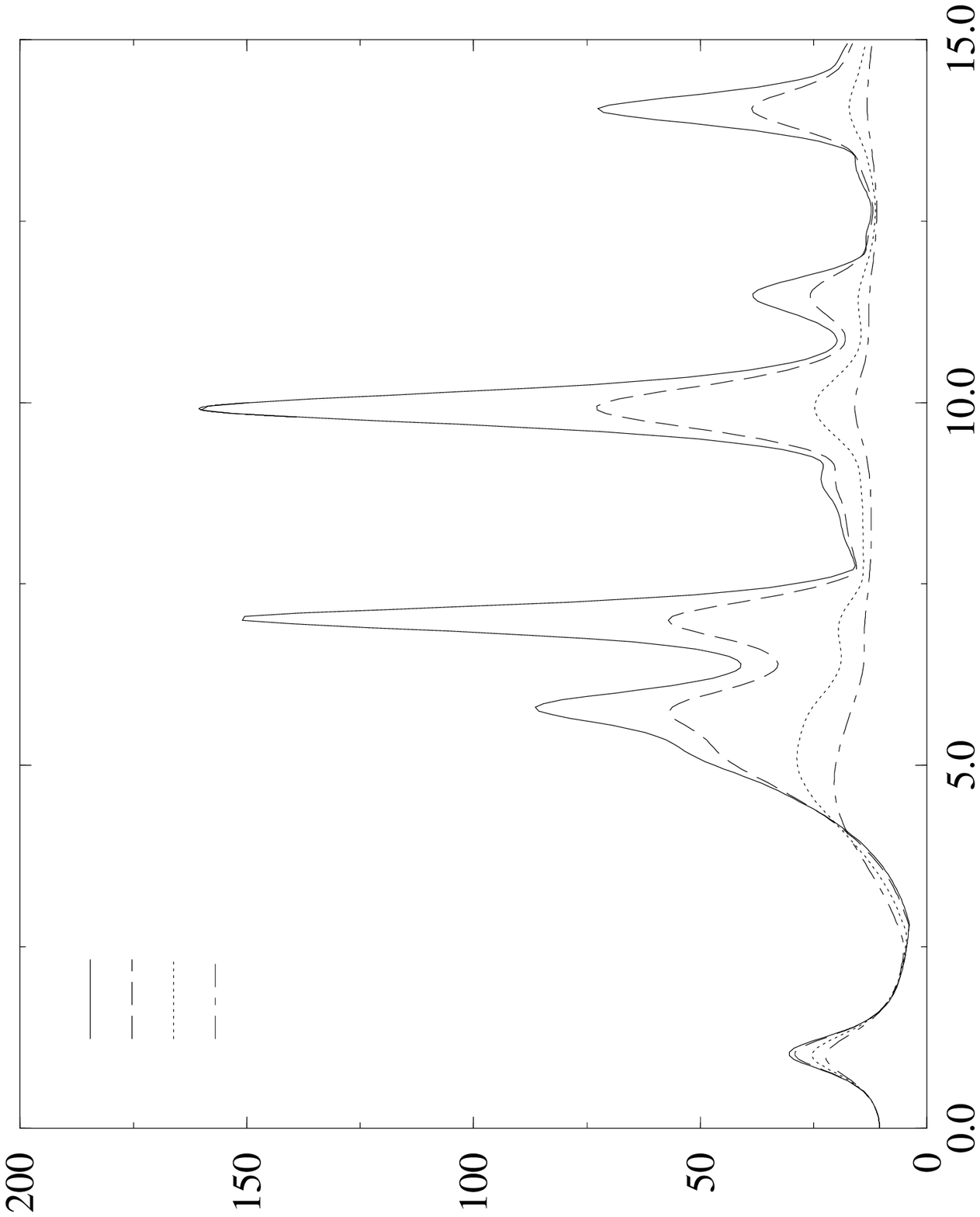}
\put(80,0){\small$\theta$}
\put(9,53){\small$\xi$}
\put(43,82){\small$n_b=0.0$}
\put(43,78){\small$n_b=0.1$}
\put(43,74){\small$n_b=1.0$}
\put(43,70){\small$n_b=10.0$}
\put(30,60){\small$N=10.0$}
\end{picture}
\figcap{Correlation lengths for different values of $n_b$.
The high peaks occur for trapping states and go away as
$n_b$ increases.}
\label{f-xinb}
\end{figure}

In order to see whether trapping states influence  the  correlation  length  we
pre\-sent
in \fig{f-FTFxid} a similar  spectral  decomposition  of  $\xi(\theta)$
(lower graph) and we find the same peaks. A more direct way of seeing  the
effect of trapping states is to study the correlation length for  small  $n_b$.
In \fig{f-xinb} we see some very pronounced peaks for small $n_b$ which rapidly
go away when $n_b$ increases. They are located at $\theta=\pi k \sqrt{N/n}$ for
every integer $k$ and $n$. The effect is most dramatic when $k$  is  small.  In
\fig{f-xinb}  there  are  conspicuous  peaks  at  $\theta=\pi  \sqrt{10}  \cdot
\{1/\sqrt{3}, ~1/\sqrt{2}, ~1, ~2/\sqrt{3}, ~2/\sqrt{2}\}$, agreeing well with
the formula for  trapping  states.
Notice how sensitive the correlation length is to the temperature when $n_b$ is
small \cite{Meystre88}.


\section{Conclusions}
\label{conclusions}
\seqnoll

We have thoroughly discussed various aspects of long-time correlations  in  the
micromaser. It is truly remarkable that this simple dynamical system  can  show
such a rich structure of different phases. The  two  basic  parameters  in  the
theory are the time the atom spends in  the  cavity,  $\tau$,  and  the  ratio
$N=R/\gamma$ between the rate at which atoms arrive and the decay constant  of
the cavity. The natural observables  are  related  to  the  statistics  of  the
outgoing atom beam, the average excitation being the simplest one. We  propose
to use the long-time correlation  length  as  a  second  observable  describing
different aspects of the photon statistics in the cavity. The  phase  structure
we have investigated is defined  in  the  limit  of  large  flux,  and  can  be
summarized as follows:

\begin{itemize}

\item
{\bf Thermal phase, $0\simleq\theta<1$.}\\
The mean number of photons $\<n\>$
is  low  (finite  in  the  limit  $N\goto\infty$),  and  so  is  the  variance
$\sigma_n$ and the correlation length $\xi$.

\item
{\bf Transition to maser phase, $\theta\simeq 1$.}\\
The maser is starting to get pumped up and $\xi$, $\<n\>$, and $\sigma_n$  grow
like $\sqrt{N}$.

\item
{\bf Maser phase, $1<\theta<\theta_1\simeq4.603$.}\\ The maser is pumped up  to
$\<n\>\sim N$, but fluctuations remain  smaller,  $\sigma_n\sim  \sqrt{N}$,
whereas
$\xi$ is finite.

\item
{\bf  First  critical  phase,  $\theta_1<\theta<\theta_2\simeq  7.790$.}\\  The
correlation length increases exponentially  with  $N$, but  nothing  particular
happens with $\<n\>$ and $\sigma_n$ at $\theta_1$.

\item
{\bf Second maser transition, $\theta\simeq 6.6$}\\ As the  correlation  length
reaches its maximum, $\<n\>$ makes a discontinuous  jump  to  a  higher  value,
though in both phases it is of the order of $N$. The fluctuations grow like $N$
at this critical point.
\end{itemize}

At higher values of $\theta$ there  are  more  maser  transitions  in  $\<n\>$,
accompanied by critical growth of $\sigma_n$, each time the photon distribution
has two competing maxima. The correlation length remains exponentially large as
a function of $N$, as long as there are several maxima, though the  exponential
factor depends on the details of the photon distribution.

No quantum interference effects have been important in  our  analysis  and  the
statistical aspects are purely classical. The reason is that we only study  one
atomic observable, the excitation level, which  can  take  the  values  $\pm1$.
Making an analogy with a spin system,
 we can say that we only measure  the  spin
along one direction. It would be  very  interesting  to  measure  non-commuting
variables, \ie\ the spin in different directions or linear superpositions of an
excited and decayed atom, and see how the phase transitions can be described in
terms of such observables \cite{Krause86,Zaheer89}. Most effective
descriptions
of phase transitions in quantum field theory rely on classical concepts,  such
as the free energy and the expectation value of some field, and do not describe
coherent effects. Since linear superpositions of excited and decayed atoms  can
be injected into the cavity,  it  therefore  seems  to  be  possible  to  study
coherent phenomena in phase transitions both theoretically and  experimentally,
using the micromaser.


\section*{Acknowledgements}
\seqnoll

B.~L. and B.-S.~S. wish to thank Gabriele Veneziano and the TH Division for the
hospitality at CERN  when  this  work  was  carried  out  and  I.~Lindgren  for
providing early guidance to the experimental work. The research by B.-S.~S. was
supported in part by the Swedish National Research Council under  contract  No.
8244-316, in part  by  the  Research  Council  of  Norway  under  contract  No.
420.95/004. The research of B.~L. was supported in part by the  Danish
Research
Councils  for  the  Natural  and  Technical   Sciences   through   the   Danish
Computational  Neural  Network  Center  ({\sc  connect})  under  contracts  No.
5.21.08.07 and 5.26.18.18.

\newpage
\appendix


\section{Jaynes--Cummings with damping}
\seqnoll
\label{AppDamping}

In most experimental situations the time the atom spends in the cavity is small
compared to the average time between the  atoms  and  the  decay  time  of  the
cavity. Then it is a good  approximation  to  neglect  the  damping  term  when
calculating the  transition probabilities from the cavity--atom  interaction.
In order to establish the range of validity of the approximation we  shall  now
study the full interaction governed by the JC Hamiltonian in \eq{JCH}  and  the
damping in \eq{Damping}. The density matrix for the cavity and one atom can  be
written as

\begin{formula}{dmca}
   \rho=\rho^0\otimes\id +\rho^z\otimes\sigma_z
   +\rho^+\otimes\sigma\_+\rho^-\otimes\sigma_+~~,
\end{formula}

\noi   where   $\rho^\pm=\rho^x\pm   i\rho^y$   and    $\sigma_\pm=(\sigma_x\pm
i\sigma_y)/2$. We want to restrict the cavity part of the density matrix to  be
diagonal, at least the $\rho_0$ part,
 which is the only part of  importance  for
the following atoms, provided that  the  first  one  is  left  unobserved  (see
discussion in Section~\ref{mixed}). Introducing the notation

\begin{eqnarray}\label{rhonot}
   \rho^0_n&=&\< n|\rho_0|n\>~~,\nn
   \rho^z_n&=&\< n|\rho_z|n\>~~,\\
   \rho^\pm_n&=&\< n|\rho_+|n-1\>-
   \< n-1|\rho_-|n\>~~,\non
\end{eqnarray}
\noi the equations of motion can be written as

\begin{eqnarray}\label{rhoeqom}
    \frac{d\rho^0_n}{dt}&=&
    \frac{ig}{2}(\sqrt{n}\rho^\pm_n-\sqrt{n+1}\rho^\pm_{n+1})
    -\gamma \sum_m L^C_{nm}\rho^0_m~~,\nn
    \frac{d\rho^z_n}{dt}&=&
    -\frac{ig}{2}(\sqrt{n}\rho^\pm_n+\sqrt{n+1}\rho^\pm_{n+1})
    -\gamma\sum_m L^C_{nm}\rho^z_m~~,\\
    \frac{d\rho^\pm_n}{dt}&=&
    i2g\sqrt{n}(\rho^0_n-\rho^0_{n-1}-\rho^z_n-\rho^z_{n-1})
    -\gamma \sum_m L^\pm_{nm}\rho^\pm_m~~,\non
\end{eqnarray}

\noi where

\begin{formulas}{LCLpm}
   L^C_{nm}&=&[(n_b+1)n+n_b(n+1)]\,\delta_{n,m}
           -(n_b+1)(n+1)\,\delta_{n,m-1}
           -n_b n \,\delta_{n,m+1}~~,\nn
   L^\pm_{mn}&=&[n_b(2n+1)-{\textstyle\inv{2}}]\,\delta_{n,m}
           -(n_b+1)\sqrt{n(n+1)}\,\delta_{n,m-1}
           -n_b\sqrt{n(n-1)}\,\delta_{n,m+1}~~.
\end{formulas}

\noi
It is thus consistent to study the particular form  of  the  cavity  density
matrix,
which has only one non-zero diagonal or subdiagonal for each  component,
even when damping is included. Our strategy shall be  to  calculate  the
first-order
correction  in  $\gamma$  in  the   interaction   picture,   using   the
JC Hamiltonian as the free part. The JC part of \eq{rhoeqom} can be drastically
simplified using the variables

\begin{eqnarray}\label{newrho}
    \rho_n^s&=&\rho^n_0+\rho^{n-1}_0-\rho_z^n+\rho_z^{n-1}~~,\nn
    \rho_n^a&=&\rho^n_0-\rho^{n-1}_0-\rho_z^n-\rho_z^{n-1}~~.
\end{eqnarray}

\noi
The equations of motion then take the form

\begin{eqnarray}\label{newreq}
   \frac{d\rho^s_n}{dt}&=&-\frac{\gamma}{2}\sum_m
               \left[(L^C_{nm}+L^C_{n-1,m-1})\rho^s_m+
               (L^C_{nm}-L^C_{n-1,m-1})\rho^a_m\right]~~,\nn
   \frac{d\rho_a^n}{dt}&=&2ig\sqrt{n}\rho^\pm_{n}-\frac{\gamma}{2} \sum_m
               \left[(L^C_{nm}-L^C_{n-1,m-1})\rho^s_m+
                            (L^C_{nm}+L^C_{n-1,m-1})\rho^a_m\right],\\
   \frac{d\rho^\pm_n}{dt}&=&2ig\sqrt{n} \rho^a_n
           -\gamma \sum_m L^\pm_{nm}\rho^\pm_m~~.\non
\end{eqnarray}

\noi
The  initial  conditions  $\rho^s_n(0)=p_{n-1}$,   $\rho^a_n(0)=-p_{n-1}$   and
$\rho^\pm_n(0)=0$ are obtained from

\begin{eqnarray}\label{beg}
    \Tr\(\rho(0) |n\>\< n|\otimes\id\)&=&2 \rho_0^n(0)=p_n~~,\nn
    \Tr\(\rho(0) |n\>\< n|\otimes\inv{2}(\id-\sigma_z)\)&=&
    \rho_0^n(0)-\rho_z^n(0)=0~~,\\
    \Tr\(\rho(0) |n\>\< n|\otimes\sigma_x\)&=&
    \Tr\(\rho(0) |n\>\< n|\otimes\sigma_y\)=0~~.\non
\end{eqnarray}

\noi
In the limit $\gamma\goto 0$ it is easy to solve \eq{newreq} and  we  get  back
the standard solution of the JC equations, which is

\begin{eqnarray}\label{JCsol}
   \rho_s^n(t)&=&p_{n-1}~~,\nn
   \rho_a^n(t)&=&-p_{n-1} \cos(2gt\sqrt{n})~~,\\
   \rho_\pm^n(t)&=&-i p_{n-1}  \sin(2gt\sqrt{n})~~.\non
\end{eqnarray}

\noi
Equation   (\ref{newreq})    is    a    matrix    equation    of    the    form
$\dot{\rho}=(C_0-\gamma C_1)\rho$. When $C_0$ and $C_1$ commute the solution
can
be written as $\rho(t)=\exp(\gamma C_1  t)\exp(C_0  t)\rho(0)$,  which  is  the
expression used in \eq{stat1}. In our case $C_0$ and $C_1$ do not  commute  and
we have to solve the equations perturbatively in $\gamma$.  Let  us  write  the
solution  as  $\rho(t)=\exp(C_0  t)\rho_1(t)$  since  $\exp(C_0  t)$   can   be
calculated explicitly. The equation for $\rho_1(t)$ becomes

\begin{formula}{r1eq}
    \frac{d\rho_1}{dt}=-\gamma e^{-C_0 t}C_1e^{C_0t}\rho_1(t)~~,
\end{formula}

\noi
which to lowest order in $\gamma$ can be integrated as

\begin{formula}{r1int}
    \rho_1(\tau)=-\gamma \int_0^\tau dt\, e^{-C_0 t}C_1e^{C_0t}\rho(0)
    +\rho(0)~~.
\end{formula}

\noi
The explicit expression for $\exp( C_0 t)$ is

\begin{formula}{C0}
        e^{C_0 t}=\delta_{nm}
        \(\begin{array}{ccc} 1 & 0 & 0\\
        0 & \cos(2gt\sqrt{n}) & i\sin(2gt\sqrt{n}) \\
        0 & i \sin(2gt\sqrt{n}) & \cos(2gt\sqrt{n}) \end{array}\)~~,
\end{formula}

\noi
and, therefore, $\exp(-C_0t)C_1\exp(C_0t)$ is a bounded  function  of  $t$.
The
elements  of  $C_1$  are  given  by  various  combinations  of  $L^C_{nm}$  and
$L^\pm_{nm}$ in \eq{LCLpm} and they grow  at  most  linearly  with  the  photon
number. Thus the integrand of \eq{r1int} is of the order of  $\<n\>$  up  to an
$n_b$-dependent factor. We conclude that the damping is negligible as  long  as
$\gamma\tau\<n\>\ll 1$,
unless $n_b$ is very large. When  the  cavity  is  in a
maser phase, $\<n\>$  is  of  the  same  order  of  magnitude  as
$N=R/\gamma$, so the condition becomes $\tau R\ll 1$.

\bort{
Even though this equation can be integrated explicitly it  only  results  in  a
very complicated expression which does not tell us directly anything about  the
approximation. It is more useful to estimate the size by recognizing that $C_0$
only has one real eigenvalue which is equal to zero, and two imaginary ones, so
the norm of $\exp(-C_0 t)$ is 1. The condition for neglecting the damping while
the atom is in the cavity is that $\gamma C_1 \tau  \rho_1$  should  be  small.
Since $C_1$ is essentially linear in $n$ the condition reads  $\gamma  \tau  \<
n\> \ll 1$. In the maser phase and above we have that $\< n\>$ is of  the  same
order of magnitude as $N=R/\gamma$, so the condition becomes $\tau R\ll 1$.
}

\section{Sum rule for the correlation lengths}\label{AppSumRule}
\seqnoll

In this appendix we derive the sum rule quoted in Eq.~(\ref{SubSumRule}) and
use the notation of Section \ref{EigenvalueProblem}.

For $A_K=0$ the determinant $\det L_K$ becomes $B_0B_1\cdots  B_K$  as  may  be
easily derived by row manipulation. Since $A_K$ only occurs  linearly  in  the
determinant it must obey the recursion relation $\det  L_K=B_0\cdots  B_K  +A_K
\det L_{K-1}$. Repeated application of this relation leads to the expression

\begin{formula}{Determinant}
\det L_K=\sum_{k=0}^{K+1} B_0\cdots B_{k-1}A_k\cdots A_K~~.
\end{formula}

\noi This is valid for arbitrary values of $B_0$ and $A_K$. Notice that here we
define $B_0\cdots B_{k-1}$ $=1$
for $k=0$  and  similarly  $A_k\cdots  A_K=1$  for
$k=K+1$.

In the actual case  we  have  $B_0=A_K=0$,  so  that  the  determinant
vanishes. The characteristic polynomial consequently takes the form

\begin{formula}{}
\det(L_K-\lambda)=(-\lambda)   (\lambda_1-\lambda)   \cdots
(\lambda_K- \lambda)=-D_1\lambda+D_2\lambda^2+\O(\lambda^3)~~,
\end{formula}

\noi where the last expression is valid for $\lambda\to0$. The coefficients
are

\begin{formula}{}
D_1=\lambda_1\cdots\lambda_K~~,
\end{formula}

\noi and

\begin{formula}{}
D_2=D_1\sum_{k=1}^K \frac1{\lambda_k}~~.
\end{formula}

To calculate $D_1$ we note that it is the sum of the $K$ subdeterminants  along
the diagonal. The subdeterminant obtained by removing the $k$'th row and column
takes the form

\begin{formula}{}
\left|\begin{array}{cccccccc}
A_0+B_0&-B_1&\\
&&\vdots\\
&-A_{k-2}&A_{k-1}+B_{k-1}&0\\
&&0&A_{k+1}+B_{k+1}&-B_{k+2}\\
&&&\vdots\\
&&&&-A_{K-1}&A_K+B_K\\
\end{array}\right|~~,
\end{formula}

\noi which decomposes into the product of two  smaller  determinants which
may be calculated using Eq.~(\ref{Determinant}). Using that $B_0=A_K=0$ we get

\begin{formula}{}
D_1=\sum_{k=0}^K A_0\cdots A_{k-1} B_{k+1}\cdots B_K~~.
\end{formula}

\noi Repeating this procedure for $D_2$ which is a sum of all possible diagonal
subdeterminants with two rows and columns removed ($0\le k<l\le K$) , we find

\begin{formula}{}
D_2=\sum_{k=0}^{K-1}\sum_{l=k+1}^K \sum_{m=k+1}^l
A_0\cdots A_{k-1} B_{k+1}\cdots B_{m-1}A_m\cdots A_{l-1}B_{l+1}\cdots B_K~~.
\end{formula}

\noi Finally, making use of Eq.~(\ref{EquilAB}) we find

\begin{formula}{}
D_1=\frac{B_1\cdots B_K}{p^0_0}\sum_{k=0}^K p^0_k~~,
\end{formula}

\noi and

\begin{formula}{}
D_2=\frac{B_1\cdots B_K}{p^0_0}
\sum_{k=0}^{K-1}\sum_{l=k+1}^K \sum_{m=k+1}^l \frac{p^0_kp^0_l}{B_mp^0_m}~~.
\end{formula}

\noi Introducing the cumulative probability

\begin{formula}{}
P^0_n=\sum_{m=0}^{n-1} p^0_m~~,
\end{formula}

\noi and interchanging the sums, we get the correlation sum rule

\begin{formula}{SumRule}
\sum_{n=1}^K
\frac1{\lambda_n}=\sum_{n=1}^K\frac{P^0_n(1-P^0_n/P^0_{K+1})}{B_np^0_n}~~.
\end{formula}

This sum rule is valid for finite $K$ but diverges  for  $K\to\infty$,  because
the equilibrium distribution $p^0_n$  approaches  a  thermal  distribution  for
$n\gg N$. Hence the right-hand side diverges logarithmically in that limit. The
left-hand side also diverges logarithmically with the truncation  size  because
we have $\lambda^0_n=n$ for the untruncated thermal  distribution.  We  do  not
know the thermal eigenvalues for the truncated case, but expect that  they
will
be of the form $\lambda^0_n=n+\O(n^2/K)$ since they should vanish for $n=0$ and
become progressively worse as $n$ approaches $K$. Such a correction leads to  a
finite correction to $\sum_n 1/\lambda_n$. In fact, evaluating  the
right-hand
side of Eq.~(\ref{SumRule}), we get for large $K$

\begin{formula}{}
\sum_{n=1}^K \frac1{\lambda_n^0}
\simeq\sum_{n=1}^K \frac{1-[n_b/(1+n_b)]^n}{n}
\simeq\sum_{n=1}^K \frac1n -\log(1+n_b)~~.
\end{formula}

Subtracting the thermal case  from  Eq.~(\ref{SumRule})
we  get  in  the  limit
of
$K\to\infty$

\begin{formula}{}
\sum_{n=1}^\infty \(\frac1{\lambda_n}-\frac1{\lambda_n^0}\)
=\sum_{n=1}^\infty\(\frac{P^0_n(1-P^0_n)}{B_np^0_n}
-\frac{1-[n_b/(1+n_b)]^n}n\)~~.
\end{formula}

\noi Here we have extended the summation to infinity under the assumption  that
for large $n$ we have $\lambda_n\simeq\lambda^0_n$. The left-hand side  can  be
approximated by $\xi-1$ in regions where the leading correlation length is much
greater than the others. A comparison of the exact eigenvalue and the  sum-rule
prediction is made in Fig. \ref{FigSumRule}.

\section{Damping matrix}
\label{dampingmatrix}
\seqnoll

In this appendix we find an integral representation for the matrix elements  of
$(x+L_C)^{-1}$, where $L_C$ is given by Eq.~(\ref{LC}). Let

\begin{equation}
v_n = \sum_{m=0}^{\infty} (x\delta _{nm} +( L_{C})_{nm})w_m~~~,
\end{equation}

\noi and introduce generating functionals $v(z)$ and  $w(z)$  for complex $z$
defined by

\begin{equation}
v(z) = \sum_{n=0}^{\infty} z^n v_n~~~,\quad
w(z) = \sum_{n=0}^{\infty} z^n w_n~~~.
\end{equation}

\noi By making use of

\begin{equation}
v(z) = \sum_{n,m=0}^{\infty} (x+L_C )_{nm}z^n w_m~~~,
\end{equation}

\noi one can derive a first-order differential equation for $w(z)$,

\begin{equation}
(x+n_{b}(1-z))w(z) + (1+n_{b}(1-z))(z-1)\frac{dw(z)}{dz}=v(z)~~~,
\end{equation}

\noi which can  be  solved  with  the  initial  condition  $v(1)=1$, \ie\
$w(1)=1/x$. If we consider the monomial $v(z) = v_mz^m$ and  the  corresponding
$w(z)=w_{m}(z)$, we find that

\begin{equation}
\label{matrixelements}
w_m (z) = \int_{0}^{1} dt(1-t)^{x-1}\frac{[z(1-t(1+n_{b}))
+t(1+n_{b})]^m}{[1+n_{b}t(1 -z)]^{m+1}}~~~.
\end{equation}

\noi Therefore $(x+L_{c})^{-1}_{nm}$ is given by the coefficient of $z^n$ in
the
series expansion of $w_{m}(z)$. In particular, we obtain for $n_b = 0$ the
result

\begin{equation}
(x+L_{C})^{-1}_{nm} = \left(
\begin{array}{c} m\\n \end{array}
\right)
\frac{\Gamma (x+n)\Gamma (m-n+1)}{\Gamma (x+m+1)}~~~,
\end{equation}

\noi where $m\geq n$. We then find that

\begin{equation}
\label{above}
\P_0(+,+) = \sum_{n=0}^{\infty} \cos^{2}(g\tau\sqrt{n+1})\sum_{m=n}^{\infty}
\frac{m!}{n!}\frac{N\Gamma (N+n)}{\Gamma
(N+m+1)}\cos^{2}(g\tau\sqrt{m+1})p_{m}^{0}~~~,
\end{equation}

\noi  where   $p_{m}^{0}$   is   the   equilibrium   distribution   given   by
\eq{Equilibrium}, and where $x=N=R/\gamma $. Equation
(\ref{above})  can  also  be
derived
from  the  known  solution  of  the  master  equation
in Eq.~(\ref{Damping})  for
$n_b   =   0$
\cite{Agarwal73}. For small $n_b$ and/or large $x$, \eq{matrixelements}  can
be
used to a find a series expansion in $n_b$.

\newpage

\def\pub#1#2#3#4#5{\bibitem{#2}#3, {\sl #4}, #5.}

\newpage

\bort{
\begin{figure}[htb]
\unitlength=1mm
\begin{picture}(140,90)(0,0)
\includegraphics{lambda.eps}
\put(10,47){\small$\lambda$}
\put(30,80){\small\parbox{30mm}{$N=10$\\$n_b=0.15$}}
\put(80,3){\small$\theta$}
\end{picture}
\figcap{Mean field solution for the subleading eigenvalue.}
\label{FigMeanLambda}
\end{figure}
}

\begin{thebibliography}{xx}

\pub{Goy83}{Goy83}
{P. Goy, J.~M. Raimond, M. Gross and S. Haroche}
{Observation of cavity-enhanced single-atom spontaneous emission}
{Phys. Rev. Lett. {\bf 50} (1983) 1903}

\pub{Meschede\&al85}{Meschede85}
{D. Meschede, H. Walther and G. M\"uller}
{One-atom maser}
{Phys. Rev. Lett. {\bf 54} (1985) 551}

\pub{Walther88}{Walther88}
{H. Walther}
{The single atom maser and the quantum electrodynamics in a cavity}
{Physica Scripta {\bf T23} (1988) 165}

\pub{Walther88}{Walther88a}
{H. Walther}
{Experiments on cavity quantum electrodynamics}
{Phys. Rep. {\bf 219} (1992) 263}

\pub{Walther95}{Walther95}
{H. Walther}
{Experiments with single atoms in cavities and traps}
{in {\it Fundamental problems in quantum theory}, Eds. D. M. Greenberger
and A. Zeiler, Ann. N.Y. Acad. Sci. {\bf 755} (1995) 133}

\pub{An\&al94}{An94}
{K. An, J.~J. Childs, R.~R. Dasari, and M.~S. Feld}
{Microlaser: A laser with one atom in an optical resonator}
{Phys. Rev. Lett. {\bf 73} (1994) 3375}

\pub{haroche92}{Haroche92}
{S. Haroche}
{Cavity quantum electrodynamics}
{in {\it Fundamental systems in quantum optics}, p. 767,
Eds. J. Dalibard, J. M. Raimond and J. Zinn-Justin, (Elsevier, 1992)}

\pub{meschede92}{Meschede92}
{D. Meschede}
{Radiating atoms in confined space: From spontaneous emission to micromasers}
{Phys. Rep. {\bf 211} (1992) 201}

\pub{meystre92}{Meystre92}
{P. Meystre}
{Cavity quantum optics and the quantum measurement process}
{{\it Progress in Optics} {\bf XXX}, p.~261, Ed. E. Wolf, (Elsevier, 1992)}

\pub{Aliskenderov\&al}{Aliskenderov93}
{E.~I. Aliskenderov, A.~S. Shumovsii and H. Trung Dung}
{Quantum effects in the interaction of an atom with radiation}
{Phys. Part. Nucl. {\bf 24} (1993) 177}

\pub{Oraevskii94}{Oraevskii94}
{A.~N. Oraevskii}
{Spontaneous emission in a cavity}
{Physics Uspekhi {\bf37} (1994) 393}

\pub{Barnet\&al86}{Barnett86}
{S.~M. Barnett, F. Filipowicz, J. Javanainen, P.~L. Knight and P.
Meystre}{The Jaynes-Cummings model and beyond}
{in {\it Frontiers in quantum optics},
Eds. E. R. Pike and S. Sarkar, (Adam Bilger, 1986)}

\pub{Milonni\&Singh}{Milonni91}
{P.~W. Milonni and S. Singh}
{Some recent developments in the fundamental theory of light}
{in {\it Advances in atomic, molecular, and optical physics} {\bf28} (1991) 75,
Eds. D. Bates and B. Bederson, (Academic Press, 1991)}

\pub{Jaynes\&Cummings63}{Jaynes63}
{E.~T. Jaynes and F.~W. Cummings}
{Comparison of quantum and semiclassical radiation theories with application to
the beam maser}
{Proc. IEEE {\bf 51} (1963)  89}

\pub{Meystre74\&al}{Meystre74}
{P. Meystre, E. Geneux, A. Quattropani and A. Faist}
{Long-time behaviour of a two-level system in interaction with an
electromagnetic
field}
{Nuovo Cimento {\bf 25B} (1975) 521}

\pub{Eberly80}{Eberly80}
{J.~H. Eberly, N.~B. Narozhny and J.~J. Sanchez-Mondragon}
{Periodic spontaneous collapse and revival in a simple quantum model}
{Phys. Rev. Lett. {\bf 44} (1980) 1323}

\pub{Narozhny80\&al}{Narozhny80}
{N.~B. Narozhny, J.~J. Sanchez-Mondragon and J.~H. Eberly}
{Coherence versus incoherence: Collapse and revival in a simple quantum model}
{Phys. Rev. {\bf A23} (1981) 236}

\pub{Knight\&Radmore82}{Knight82}
{P.~L. Knight and P.~M. Radmore}
{Quantum revivals of a two-level system driven by chaotic radiation}
{Phys. Lett. {\bf 90A} (1982) 342}

\pub{Filipowicz86b}{Filipowicz86b}
{P. Filipowicz}
{Quantum revivals in the Jaynes-Cummings model}
{J. Phys. A: Math. Gen. {\bf 19} (1986) 3785}

\pub{Arroyo\&al90}{Arroyo90}
{G. Arroyo-Correa and J.~J. Sanchez-Mondragon}
{The Jaynes-Cummings model thermal revivals}
{Quantum Opt. {\bf 2} (1990) 409}

\pub{Rempe\&al87}{Rempe87}
{G. Rempe, H. Walther and N. Klein}
{Observation of quantum collapse and revival in a one-atom maser}
{Phys. Rev. Lett. {\bf 58} (1987) 353}

\pub{HBT56}{HBT56}
{R. Hanbury-Brown and R.~Q. Twiss}
{Correlation between photons in two coherent beams of light}
{Nature {\bf 177} (1956) 27}

\pub{Glauber63}{Glauber63}
{R. J. Glauber}
{The quantum theory of optical coherence}
{Phys. Rev. {\bf 130} (1963) 2529}

\pub{Boal\&al90}{Boal90}
{D.~H. Boal, C.-K. Gelbke and B.~K. Jennings}
{Intensity interferometry in subatomic physics}
{Rev. Mod. Phys. {\bf 62} (1990) 553}

\pub{Skagerstam94}{Skagerstam94}
{B.-S. Skagerstam}
{Coherent states --- Some applications in quantum field theory and particle
physics}
{in {\it Coherent states: Past, present, and the future}, p. 469,
Eds. D.~H. Feng, J.~R. Klauder and M.~R. Strayer, (World Scientific, 1994)}

\pub{Elmforsetal95}{Elmforsetal95}
{P. Elmfors, B. Lautrup and B.-S. Skagerstam}
{Correlations as a handle on the quantum state of the micromaser}
{CERN/TH 95-154 preprint, CERN, 1995}

\pub{Filipowicz\&al86}{Filipowicz86}
{D. Filipowicz, J. Javanainen and P. Meystre}
{The microscopic maser}
{Opt. Commun. {\bf 58} (1986) 327}

\pub{Filipowicz\&al86}{Filipowicz86a}
{D. Filipowicz, J. Javanainen and P. Meystre}
{Theory of a microscopic maser}
{Phys. Rev. {\bf A34} (1986) 3077}

\pub{Rempe\&Walther90}{Rempe90a}
{G. Rempe and H. Walter}
{Sub-Poissonian atomic statistics in a micromaser}
{Phys. Rev. {\bf A42} (1990) 1650}

\pub{Paul\&Richter91}
{PaulRichter91}
{H. Paul and Th. Richter}
{Bunching and antibunching of de-excited atoms leaving a micromaser}
{Opt. Commun. {\bf 85} (1991) 508}

\pub{Rempe\&al90}{Rempe90}
{G. Rempe, F. Schmidt-Kaler and H. Walther}
{Observation of sub-Poissonian photon statistics in a micromaser}
{Phys. Rev. Lett. {\bf 64} (1990) 2783}

\pub{Brigeletal94}{Brigeletal94}
{H.-J. Briegel, B.-G. Englert, N. Sterpi and H. Walther}
{One-atom maser: statistics of detector clicks}
{Phys. Rev. {\bf 49} (1994) 2962}

\pub{Knight\&Milonni80}{Knight80}
{P.~L. Knight and P.~W. Milonni}
{The Rabi frequency in optical spectra}
{Phys. Rep. {\bf66} (1980) 21}

\pub{Averbukh\&Perelman89a}{Averbukh89a}
{I.~Sh. Averbukh and N.~F. Perelman}
{Fractional regenerations of wave packets in the course of long-term
evolution of highly excited quantum systems}
{Sov. Phys. JETP {\bf 69} (1989) 464}

\pub{Averbukh\&Perelman89b}{Averbukh89b}
{I.~Sh. Averbukh and N.~F. Perelman}
{Fractional revivals: Universality in the long-term evolution of quantum
wave packets beyond the correspondence principle dynamics}
{Phys. Lett. {\bf A139} (1989) 449}

\pub{Fleischhauer\&Schleich93}{Fleischhauer93}
{M. Fleischhauer and W. Schleich}
{Revivals made simple: Poisson summation formula as a key to the
revivals in the Jaynes-Cummings model}
{Phys. Rev. {\bf A47} (1993) 4258}

\pub{Wright\&Meystre89}{Wright89}
{E.~M. Wright and P. Meystre}
{Collapse and revival in the micromaser}
{Opt. Lett. {\bf 14} (1989) 177}

\pub{Guzman\&al89}{Guzman89}
{A.~M. Guzman, P. Meystre and E.~M. Wright}
{Semiclassical theory of the micromaser}
{Phys. Rev. {\bf A40} (1989) 2471}

\pub{Herzog95}{Herzog95}
{U. Herzog}
{Micromasers with stationary non-Poissonian pumping}
{Phys. Rev. {\bf A52} (1995) 602}

\pub{Krause\&al86}{Krause86}
{J. Krause, M. Scully and H. Walther}
{Quantum theory of the micromaser: Symmetry breaking via off-diagonal atomic
injection}
{Phys. Rev. {\bf A34}  (1986) 2032}

\pub{Lugiato\&al87}{Lugiato87}
{L. Lugiato, M. Scully and H. Walther}
{Connection between microscopic and macroscopic maser theory}
{Phys. Rev. {\bf A36} (1987)  740}

\pub{Zaheer\&Zubairy89}{Zaheer89}
{K. Zaheer and M.~S. Zubairy}
{Phase sensitivity in atom-field interaction via coherent superposition}
{Phys. Rev. {\bf A39} (1989) 2000}

\pub{Schleich\&Wheeler87}{Schleich87}
{W. Schleich and J.~A. Wheeler}
{Oscillations in photon distribution of squeezed states and interference in
phase space}
{Nature {\bf 326} (1987) 574}

\pub{Agarwal73}{Agarwal73}
{G.~S. Agarwal}
{Master equation methods in quantum optics}
{in {\it Progress in Optics} {\bf XI}, p.~1,
Ed. E. Wolf, (North Holland, 1973)}

\bibitem{Walls95}
{D.~F. Walls and G.~J. Milburn,
{\sl Quantum optics}, (Springer, 1995)}

\pub{Rempe\&al91}{Rempe91}
{G. Rempe, M.~O. Scully and H. Walther}
{The one-atom maser and the generation of non-classical light}
{Physica Scripta {\bf T34} (1991) 5}

\pub{Teich\&Saleh88}{Tei88}
{M.~C. Teich and B.~E.~A. Saleh}
{Photon bunching and antibunching}
{in {\it Progress in Optics} {\bf XXVI}, p.~1, Ed. E. Wolf, (North-Holland,
1988)}

\pub{Mandel79}{Mandel79}
{L. Mandel}
{Sub-Poissonian photon statistics in resonance fluorescence}
{Opt. Lett. {\bf 4} (1979) 205}

\bibitem{Fritz79}
{F.-J. Fritz, B. Huppert and W. Willems,
{\it Stochastische matrizen},
(Springer-Verlag, 1979).}

\pub{Meystre\&al88}{Meystre88}
{P. Meystre, G. Rempe and H. Walther}
{Very-low-temperature behavior of a micromaser}
{Opt. Lett. {\bf 13} (1988) 1078}

\pub{Filipowicz\&al86}{Filipowicz86c}
{P.~Filipowicz, J.~Javanainen and P.~Meystre}
{Quantum and semiclassical steady states of a kicked cavity mode}
{J.~Opt.~Soc.~Am. {\bf B3} (1986) 906}

\bibitem{Schuss80}
{Z. Schuss, {\it Theory and applications of
stochastic differential equations},
(John Wiley and Sons, 1980).}

\pub{BogarBH94}{BogarBH94}
{P.~Bog\'{a}r, J.~A.~Bergou and M.~Hillery}
{Quantum island states in the micromaser}
{Phys. Rev. {\bf A50} (1994) 754}

\pub{WagnerSW94}{WagnerSW94}
{C.~Wagner, A.~Schezle and H.~Walther}
{Atomic waiting-time and correlation functions}
{Opt. Commun. {\bf 107} (1994) 318}

\pub{Herzog94}{Herzog94}
{U.~Herzog}
{Statistics of photons and de-excited atoms in a micromaser
with Poissonian pumping}
{Phys. Rev. {\bf A50} (1994) 783}

\end{thebibliography}
\end{document}